\documentclass[bibyear,referee]{aa} 
\newcommand{\hv}{$H_{\rm V}$}
\newcommand{\degs}{$^{\rm o}$}

\usepackage{graphicx}
\usepackage{txfonts}
\usepackage{lscape}
\begin{document} 

\titlerunning{Absolute magnitudes of TNOs}

   \title{Absolute magnitudes and phase coefficients of trans-Neptunian objects\thanks{Based in part on observations collected
at the German-Spanish Astronomical Center, Calar Alto, operated
jointly by Max-Planck-Institut f\"ur Astronomie and Instituto de
Astrof\'isica de Andaluc\'ia (CSIC). Based in part on observations made with the Isaac Newton Telescope operated on the island of La Palma by the Isaac 
Newton Group in the Spanish Observatorio del Roque de los Muchachos of the Instituto de Astrofísica de Canarias.
Partially based on data obtained with the 1.5 m telescope, which is operated by the Instituto de Astrof\'isica de Andaluc\'ia at the Sierra Nevada Observatory.
Partially based on observations obtained at the Southern Astrophysical Research (SOAR) telescope, which is a joint project of the 
Minist\'erio da Ci\^encia, Tecnologia, e Inova\c c\~ao (MCTI) da Rep\'ublica Federativa do Brasil, the U.S. National Optical Astronomy Observatory (NOAO), 
the University of North Carolina at Chapel Hill (UNC), and Michigan State University (MSU).
Based in part on observations made at the Observat\'orio Astron\^omico do Sert\~ao de Itaparica operated by the Observat\'orio Nacional / MCTI, Brazil.
Partially based on observations made with the Liverpool Telescope operated on the island of La Palma by Liverpool John Moores University in the Spanish 
Observatorio del Roque de los Muchachos of the Instituto de Astrof\'isica de Canarias with financial support from the UK Science and Technology Facilities Council.
}}


   \author{A. Alvarez-Candal\inst{1}
\and
N. Pinilla-Alonso\inst{2}
\and
J.L. Ortiz\inst{3}
\and
R. Duffard\inst{3}
\and
N. Morales\inst{3}
\and
P. Santos-Sanz\inst{3}
\and
A. Thirouin\inst{4}
\and
J.S. Silva\inst{1} 
}

   \institute{Observat\'orio Nacional / MCTI, Rua General Jos\'e Cristino 77, Rio de Janeiro, RJ, 20921-400, Brazil\\
              \email{alvarez@on.br}
         \and
             Department of Earth and Planetary Sciences, University of Tennessee, Knoxville, TN, 37996, United States
         \and
             Instituto de Astrofísica de Andalucía, CSIC, Apt 3004, 18080, Granada, Spain
         \and
             Lowell Observatory, 1400 W Mars Hill Rd, Flagstaff, 86001 Arizona, USA
             }

   \date{Received --- --, ----; accepted --- --, ----}

 
  \abstract
   {Accurate measurements of diameters of trans-Neptunian objects are extremely complicated to obtain. Thermal modeling can provide good results, 
{\bf but accurate absolute magnitudes are needed to constrain the thermal models and derive diameters and geometric albedos.} 
The absolute magnitude, \hv, {\bf is defined as the magnitude of the object reduced to unit helio- and geocentric distances and a zero solar phase angle}
and is determined using phase curves.

Phase coefficients can also be obtained from phase curves. These are related to surface properties, yet not many are known.}
   {Our objective is to measure accurate $V$ band absolute magnitudes and phase coefficients for a sample of trans-Neptunian objects, many of which have 
been observed, and modeled, within the ``TNOs are cool'' program, one of Herschel Space Observatory key projects.}
   {We observed 56 objects using the $V$ and $R$ filters. These data, along with those available in the literature, were used to obtain phase curves and measure
$V$ band absolute magnitudes and phase coefficients by assuming a linear trend of the phase curves and considering magnitude variability due to rotational light-curve.}
   {We obtained 237 new magnitudes for the 56 objects, six of them with no reported previous measurements. Including the data from the literature
we report a total of 110 absolute magnitudes with their respective phase coefficients. The average value of \hv~is 6.39, bracketed by a minimum of
14.60 and a maximum of -1.12. In the case of the phase coefficients we report 0.10 mag per degree as the median value {\bf and a very large dispersion, ranging from
-0.88 up tp 1.35 mag per degree.}}
   {}

   \maketitle
%

\section{Introduction}

The phase curve of a minor body shows how the reduced magnitude\footnote{The observed standard magnitude normalized to the distance of the Sun and Earth.} 
of the body changes with phase angle. The phase angle, $\alpha$, is defined as the angle, measured at the location of the body, that the Earth and the Sun 
subtend. These curves show a complex behaviour, {\bf for phase angles between 5\degs and 30\degs they follow an overall linear trend}, while at small angles a departure from
linearity often occurs. In 1956 T. Gehrels coined the expression ``opposition effect'' and attributed it to the sudden increase of brightness at small
$\alpha$ shown in the phase curve of asteroid 20 Massalia (Gehrels 1956), although no explanation was offered. {\bf Since then
many works have modeled phase curves}, with or without opposition effect, analysing the relationship between these curves and
the properties of the surface: particle sizes, scattering properties, albedos, compaction, or composition, either by using astronomical or laboratory data, or
theoretical modeling (e.g., Hapke 1693, Bowell et al.1989, Nelson et al. 2000, Shkuratov et al. 2002 and references therein).

Besides providing information about surface properties, phase curves are also important because using them we can measure the absolute magnitude, $H$, 
of an airless body. $H$ is defined as the reduced magnitude of an object at $\alpha=0$\degs. 
Moreover, $H$ is related to the diameter of the body, $D$, and its geometric albedo $p$. {\bf If we are considering magnitudes in the $V$ band}, then
\begin{equation}\label{diameter}
D~[{\rm km}]=1.324\times\frac{10^{(3-H_{\rm V}/5)}}{\sqrt{p_{\rm V}}}.
\end{equation}

{\bf The first minor bodies to have their phase curves measured were asteroids (for instance the aforementioned work by Gehrels in 1956). Nowadays, we know that low-albedo
(taxonomic classes D, P, or C) asteroids show lower opposition effect spikes than higher-albedo asteroids (S or M asteroids) (Belsakya and Shevchenko \cite{belsk00}). Also
modern technologies allowed us to obtain incredible data of a handful of objects, see the recent work on comet 67P/Churyumov-Gerasimenko by Fornasier et al. (\cite{forna15}) using 
data from the ROSETTA spacecraft; or huge databases, such as the 250,000 absolute magnitudes of asteroids presented by Vere\v s et al. (\cite{veres15}) from Pan-STARRS.

Unfortunately such data is not yet available for objects farther away in the solar system, with the exception of 134340 Pluto. Therefore,}
many of the physical characteristic of the trans-Neptunian population, for instance size, albedo, or density, are still hidden from us due to the limited 
quality of the information we can currently obtain: visible and/or near-infrared spectroscopy of about 100 objects (Barucci et al. 2011, and references therein), 
colors of about 300 (Hainaut et al. 2012) drawn from a known population of more than 1,400 objects. Moreover, these data belong to the largest known 
trans-Neptunian objects, TNOs, the most easily observed ones, or some Centaurs, which is a population of dynamically unstable objects
orbiting among the giant planets, considered as coming from the trans-Neptunian region and therefore representative of this population.
Nevertheless, considerable progress has been made in understanding the 
dynamical structure of the region, but the bulk of the physical characteristics of the bodies that inhabit it remains poorly determined. 
Several observational studies conducted in the last years show a vast heterogeneity on physical and chemical properties.

With the objective of enlarging our knowledge of the TNO population, The Herschel Open Time Key Program on TNOs
and Centaurs: ``TNOs Are Cool'' (M\"uller et al. 2007) was granted with 372.7 hours of observation on the Herschel Space Observatory
(HSO). The observations are complete with a sample of 130 observed objects. 
The observed data are fed into thermal models (M\"uller et al. 2010) where a series of free parameters are fitted, among
them are $p_{\rm V}$ and $D$. These two quantities could be constrained using ground based data and thus fixing at least one of
them in the modeling improving the accuracy of the results. Among the targets observed with Herschel there are several of them that do not have 
reliable \hv~magnitude, which is fundamental for the computation of $D$ and $p_{\rm V}$ (i.e. less uncertainties in \hv~mean less uncertainties in  $D$ and 
$p_{\rm V}$).

Therefore, the HSO program ``TNOs are Cool'' needs support observations from ground-based telescopes.

One critical problem that arises when studying phase curves of TNOs is the fact that $\alpha$ can only attain low values when observing from Earth-based facilities. 
For comparison: a typical main belt asteroid can be observed up to 20\degs~or 30\degs, while a typical TNO can only reach up to 2\degs.
{\bf This means that, for TNOs, we are observing well within the opposition effect region, which prevents us to use the full power of photometric models. On the
other hand, the phase curves are very well approximated by linear functions within this restricted phase angle region (e.g., Sheppard and Jewitt \cite{schep02}).} 
Some effort has been made in this direction 
(see review by Belskaya et al. 2008, or the recent works by Perna et al. 2013 and B\"ohnhardt et al. 2014) but most of these 
use limited samples (usually one observation) assuming average values of the phase coefficients.

With this in mind we started a survey in various telescopes to obtain $V$ and $R$ magnitudes for several TNOs at as many different phase angles
as possible to measure phase curves and through them determine \hv. The survey is being carried in both hemispheres using telescopes at different locations.
In the next section we describe the observations carried and the facilities where the data were obtained. In Sect. 3 we present the results, while their analysis
is presented in Sect. 4. The discussion and some conclusions obtained from this work are presented in Sect. 5.

\section{Observations and data reduction}

The data presented in this work were collected during several observing runs spanning between September 2011 and July 2015 for well over 40 nights.
The instruments and facilities used were: {\bf The Calar Alto Faint Object Spectrograph at the 2.2-m telescope, CAHA2.2, and Multi Object Spectrograph for Calar Alto
at the 3.5-m telescope, CAHA3.5, of the Calar Alto Observatory\footnote{\tt http://www.caha.es} sited at Sierra de Los Filabres (Spain); 
the Wide Field Camera at the 2.5-m Isaac Newton Telescope, INT, sited at the Roque de los Muchachos Observatory\footnote{\tt http://www.ing.iac.es/Astronomy/telescopes/int/} 
(Spain); the direct camera at the 1.5-m telescope, OSN, of the Sierra Nevada Observatory\footnote{\tt http://www.osn.iaa.es/content/15-m-telescope} (Spain); 
the SOAR Optical Imager at the 4.1-m Southern Astrophysical Research telescope\footnote{\tt http://www.soartelescope.org/} 
sited at Cerro Pach\'on (Chile); the direct camera at the 1-m telescope of the Observat\'orio Astron\^omico do Sert\~ao de 
Itaparica\footnote{\tt http://www.on.br/impacton/}, OASI, Brazil; and the optical imaging component of the Infrared-Optical suite of instruments (IO:O) at the 2.0-m 
Liverpool telescope, Live, sited at the Roque de los Muchachos Observatory\footnote{\tt http://telescope.livjm.ac.uk/} (Spain).}
Descriptions of instruments and telescopes can be found at their respective home pages.

We always attempted to observe using the $V$ and $R$ filters sequentially, but in some cases this was not possible, either due to deterioration of weather 
conditions {\bf (i.e., no observation possible)} or instrumental/telescope problems. 
The objects were targeted, whenever possible, {\bf at different phase angles aiming at the widest spread possible.}
Along with the TNOs we targeted several standard stars fields (from Landolt 1992 and Clem and Landolt \cite{cleml13}), or were provided by the observatory, as in the
case of the Liverpool telescope, each night. 
We aimed at observing three different fields at three different airmasses per night covering the range of airmasses of our main targets.

Most observations were carried by observing the target during three exposures of 600 s per filter, although in some cases shorter exposures (300 or 400 s) 
were used to avoid saturation from nearby bright stars or trailing by faster objects (a Centaur can reach up to 2 arcsecs in 10 min). 
We did not use differential tracking. The combination of the different images allowed us to increase signal-to-noise ratio while keeping trailing 
at reasonable values. In any case, we found this approach better than, 
for instance, tracking at non-sidereal rate for 1800 s because during stacking of shorter exposures a better removal of bad pixels, cosmic ray hits, 
or background sources was obtained.

Data reduction was performed using standard photometric methods with IRAF. Master bias frames were created from daily files, as well as master flat fields 
in both filters. Files including TNOs and standard stars fields were bias and flat field calibrated. Data from the Liverpool telescope were provided already
calibrated. Identification of the targets was, for most of the 
objects, straightforward by blinking different images or, in the most complicated cases, using Aladin\footnote{\tt http://aladin.u-strasbg.fr/} (Bonnarel et al. 2000).
Instrumental apparent magnitudes were obtained using aperture photometry selecting an aperture typically three times the seeing measured in the images,
for TNOs and standard stars. Whenever a TNO was too close to another source, either by poor observing timing or by crowded fields, we performed instead 
aperture correction (see Stetson 1990).

Using the standard stars we computed extinction coefficients and color terms to correct the magnitudes of the TNOs, thus
\begin{equation}
m_0=m-\chi[k_1+k_2(v-r)],
\end{equation}
where 
$m_0$ is the apparent instrumental magnitude corrected by extinction ($v_0$ or $r_0$),
$m$ is the apparent instrumental magnitude ($v$ or $r$),
$\chi$ is the airmass, 
$k_1$ and $k_2$ are the zeroth and first order extinction coefficients, and
$(v-r)$ is the apparent instrumental color of the TNO.

Next, we translated the $m_0$ to the standard system. The transformation, to order zero, is
\begin{equation}
M=m_0+ZP,
\end{equation}
where
$M$ is the calibrated magnitude, and
$ZP$ is the zero point.
Note that, as we had many runs in the same telescopes, we computed average extinction coefficients for each site that were used whenever 
the data did not allow us to compute the night value. The same is true for $ZP$s. In the particular case of the Liverpool telescope, we used
the average extinction coefficient for the Roque de los Muchachos observatory.

{\bf Table \ref{table:obs} lists all observed objects,} along with its calibrated $V$ and $R$ magnitudes, the night the object was observed, the heliocentric
($r$) and geocentric ($\Delta$) distances and the phase angle ($\alpha$) at the moment of observation, the telescope used, and a series of notes indicating
whether we used average extinction coefficients, or average zero points, or the object had not previous data reported.

The errors in the final magnitudes include: (i) the error in the instrumental magnitudes, provided by IRAF ($\sigma_i$); (ii) the error due to 
atmospheric extintion, estimated as $\sigma_e=m_0-(m-\chi k_1)$; and the error in the calibration to the standard system, $\sigma_{ZP}$. Therefore
$\sigma^2=\sigma_i^2+\sigma_e^2+\sigma_{ZP}^2$. {\bf Note that, whenever aperture correction was performed, $\sigma_i^2=\sigma_i{_1}^2+\sigma_i{_2}^2$. Where
$\sigma_i{_1}$ is the error provided by IRAF within the smaller aperture and $\sigma_i{_2}$ is the error in the aperture correction, computed using the
task {\tt mkapfile} within IRAF.}

\section{Analysis}

In total we obtained 237 new magnitudes for 56 objects, 6 of which did not have any magnitude reported before, to the best of our knowledge.
The observed objects span {\bf from Centaurs up to Detached Objects} (from 10 to more that 100 AU)
while in eccentricity they reach values as high as 0.9. The inclinations are mostly below 40\degs~with one object at about 80\degs~and one
in retrograde orbit (2008 YB$_{3}$).

Alongside our own data we made an extensive, although not complete, search in the literature of other published $V$ and $R$ magnitudes. 
We used as our primary reference database the MBOSS 2 article by Hainaut et al. (\cite{haina12}), but we did not 
take the data directly from their catalogue. Instead we took the data from each referenced article to be included in our list.
We choose that approach because to make the phase curves we need reduced magnitudes (described in Sect. 3.2), which are computed using 
the heliocentric and geocentric distances at the moment of observation. At the same time we obtained information regarding the phase angle.
{\bf It follows that we only used data that were reported along with the site and epoch of observation.}
We obtained the orbital information from JPL-Horizons\footnote{\tt http://ssd.jpl.nasa.gov/horizons.cgi}.
In those cases where more than one magnitude was reported for the same night, for instance in light-curve analysis, we computed the average value and its 
standard deviation to use as input. In total we finished with over 1,800 individual measurements for over a hundred objects. Note that each individual 
measurement corresponds to one observing night, or entry. {\bf We did not reject any data based on their reported error bars.}

Before jumping to the results we stress three important issues: 
(i) we obtained data for 56 objects but these data alone cannot be used to create phase curves for all the objects, thus we recurred to the
literature. In the remaining of the article this augmented set of data will be called {\em our database}; (ii) as can be seen in Eq. \ref{diameter} we cannot
split albedo and diameter only using \hv, therefore whenever we speak about the {\em brightness} of an object we are referring exclusively to its magnitude
and not to its albedo properties nor its size, unless explicitly mentioned, and (iii) the magnitudes for the phase curves should be averaged over 
the rotational period to remove effect of variability due to $\Delta m>0$, which is not the case for individual measurements.

In the following subsections we describe first how we compute the colors for the complete database, and then the construction of the phase curves.

\subsection{Colors}

Being a compilation from different sources our database is very heterogeneous. Some objects have many entries, in a few cases more than fifty, while
most have less than ten entries (72\% of the sample). Not all entries have data obtained with both filters, in some occasions only $V$ filter was used, 
while in some others only the $R$ filter magnitude is available. Whenever both magnitudes were available for the same night we computed $(V-R)$. 
In this way, many objects have more than one measurement of $(V-R)$. In those cases, we computed a weighted average color which we take as 
representative for the object. By doing so we are weighting up the most precise values of $(V-R)$ instead of considering possible changes of color with 
phase angle, which is out of the scope of the present work.

\begin{figure}[ht]
\centering
\includegraphics[width=\hsize]{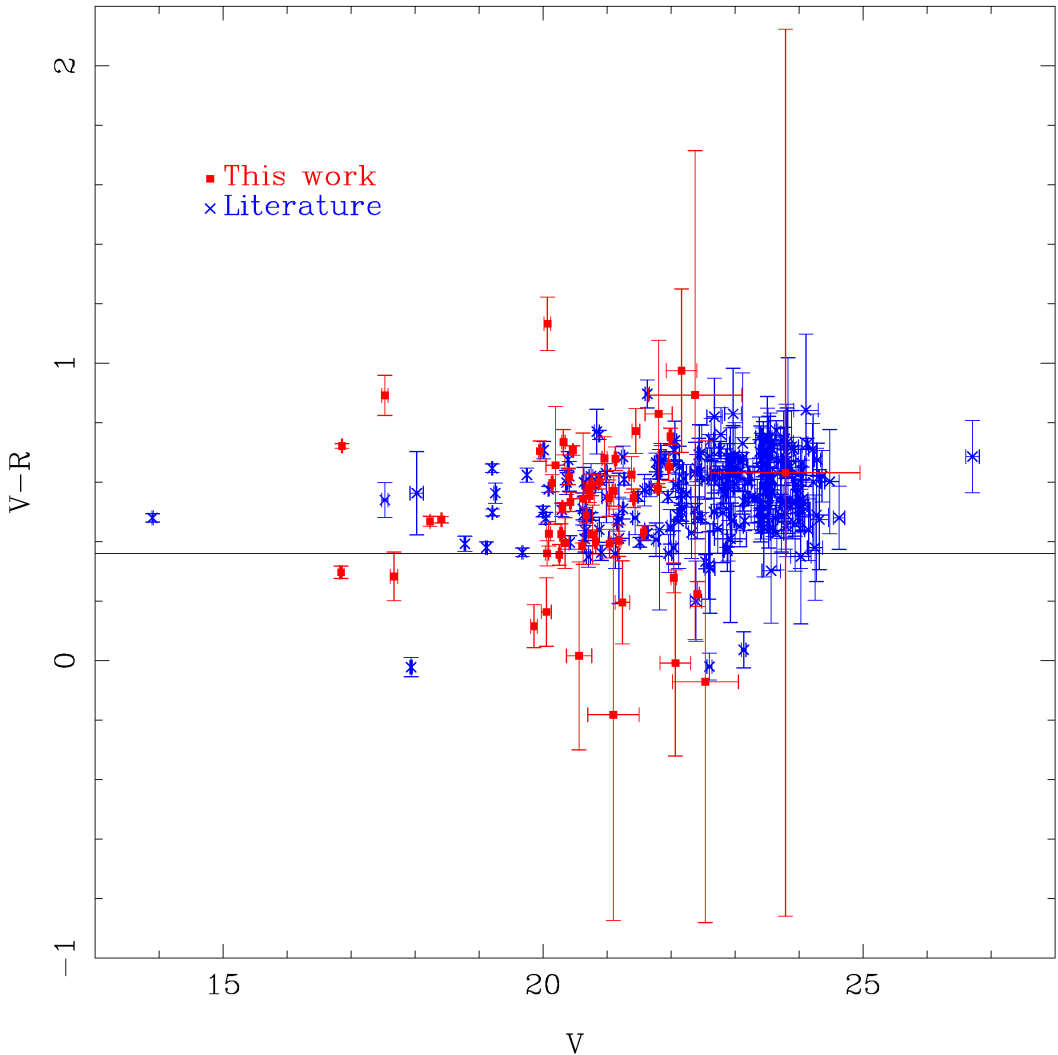}
   \caption{Color - magnitude diagram for the objects in our database. We show in red the objects that have at least one color measured by us, while
in blue appear those whose data come from the literature only. The $(V-R)_{\odot}$ is shown for reference as a horizontal line.}
      \label{colors}
\end{figure}

We show in Fig. \ref{colors} the color-magnitude diagram for all objects in our sample. 
If at least one entry for a given object was observed by us, we labeled that object as ``This work'', while if all observations for a given object were 
obtained from the literature, the label ``Literature'' was used. Note that the plot has more than 110 points. In fact we are showing the colors of objects
that do not satisfy our criteria to construct the phase curve (see below). 

Most objects shown in the figure are redder than the Sun, $(V-R)_\odot=0.36$, shown as a continuous horizontal line. Nevertheless, there are a few bluer objects, 
$(V-R)\approx0$. The great majority of the objects cluster at $V\approx23,~(V-R)\approx0.6$. From the figure it is also clear that our observations have a clear 
cut-off at about $V=22.5$, due to the size of the telescopes used, with only one object fainter than $V=23$: 2003 QA$_{91}$, having obvious large error bars.

\subsection{Phase curves}
The main objective of this work is to compute absolute magnitudes, \hv, and phase coefficients, $\beta$, of as many objects as possible. These data could
be used as complement to the Herschel Space Observatory ``TNOs are cool'' key project. Several papers have already been published presenting
\hv~of different TNOs (e.g., Sheppard and Jewitt 2002, Rabinowitz et al. 2006 and 2007, Perna et al. 2013, B\"ohnhardt et al. 2014, and others). 
We do not intend to repeat step by step these works, but to recompute the phase curves making the most of the increasing amount of data available nowadays. 
We are aware of the risks that arise due to the inhomogeneity of telescopes, instruments, detectors, and epochs. Nevertheless, we consider
important to re-analyze the available data using, if not homogeneous inputs, at least homogeneous techniques.

As mentioned above, we had to deal with the fact that not all entries (i.e., night of observation for a given object) were complete, in the sense 
that some objects for a given date were observed only in one filter, $V$ or $R$. To avoid this problem we decided to construct the individual phase curves 
using magnitudes measured with the $V$ filter. In those cases where $V$ was not available, we used the average color measured above and the $R$ magnitude
to obtain $V$. We decided, for the scope of this work, to not analize separated the $V$ and $R$ data as we are more interested in having the larger 
possible quantity of phase curves. For instance, if we use only the $V$ data, without the $R$ data, we only obtain about 50 phase curves. A similar
number of phase curves are obtained if using only $R$ data, although not necessarily the same objects.

The next step is to compute the reduced $V$ whose notation is $V(1,1,\alpha)$ which is the value used in the phase curves and
represents the magnitude of the object if located at 1 AU from the Sun and being observed at a distance of 1 AU from the Earth.

The reduced magnitude is computed from the values of $V$ and the orbital information as 
\begin{equation}
V(1,1,\alpha)=V-5\log{(r\Delta)}.
\end{equation}
We are now left with a set $\{V(1,1,\alpha),\alpha\}$ for each object.

For the phase curves we only used data for objects that were observed at three different phase angles at least. We discarded a few objects that had a small coverage 
in $\alpha$ resulting in unreliable values of \hv. We analyzed a total of 110 objects. 
For objects with no reported light-curve amplitude we assumed $\Delta m=0$ and performed a linear regression to measure \hv~via 
\begin{equation}\label{linear}
	V(1,1,\alpha)=H_{\rm V}+\alpha\times\beta,
\end{equation}
where $\beta$ is the change of magnitude per degree, also known as phase coefficient. 
Each $V(1,1,\alpha)$ was weighted by its error, assumed equal to that of the $V$ magnitude, or propagated from the $R$ magnitude and that of the average color, 
while $\alpha$ was assumed as having negligible error. By doing so we obtained \hv~as the y-intercept and $\beta$ as the slope of Eq. \ref{linear}.

We used the linear approach instead of using the full H-G system (Bowell et al. 1989) for simplicity, {\bf as we do want to add any more free parameters
that will unnecessarily complicate the interpretation of results. We also make use of the results presented in
Belskaya and Shevchenko (2000), and mentioned in the Introduction,} who showed that the opposition effect, the major departure from linearity of the phase curve, is in fact more conspicuous 
among moderate albedo objects ($p_{\rm V}>0.25$) which is not the case for most of the known TNOs 
(e.g., Lellouch et al. \cite{lello13}, Lacerda et al. \cite{lacer14}).

Some objects do have reported rotational light-curves with non-zero $\Delta m$ (we use in this work the data reported in Thirouin et al. 2010 and 2012). 
Note that $\Delta m$ can range up to half a magnitude in extreme, but rare, cases. As we are using reduced magnitudes obtained in different 
nights, and mostly individual measurements, we have to model the effect of light-curve variations on the value of $V(1,1,\alpha)$. We proceeded as follows: 
for an object with $\Delta m\neq0$
we generated from $\{V(1,1,\alpha),\alpha\}$ new sets $\{V_i(1,1,\alpha),\alpha\}$, with $i$ running from 1 to 10,000, where
\begin{equation}
V_i(1,1,\alpha)=V(1,1,\alpha)+rand_i\times\Delta m,
\end{equation}
$rand_i$ is a random number drawn from a uniform distribution within -1 and 1. By doing so, and feeding these values into Eq. \ref{linear}, 
we finish with a set $\{H_{{\rm V}i},\beta_i\}$, from where we
obtain \hv~and $\beta$ as the average over the 10,000 realizations.

\begin{figure}
   \centering
   \includegraphics[angle=0,width=\hsize]{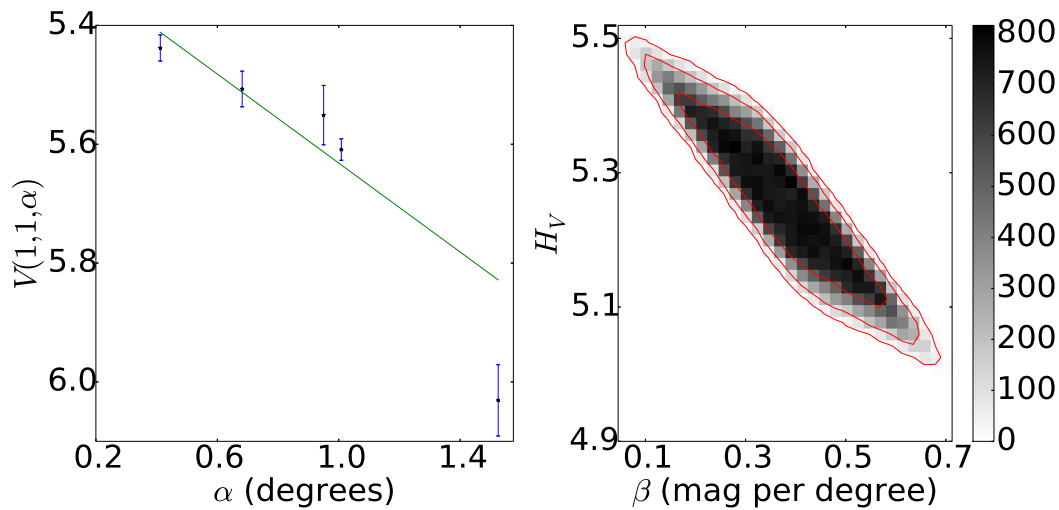}
      \caption{Example of phase curve of 1996 TL$_{66}$. 
              Left: scatter plot of $V(1,1,\alpha)$ versus $\alpha$. The line represents the solution for \hv~and $\beta$ as mentioned in the text.
              Right: density plot showing the phase space of solutions of Eq. \ref{linear} when $\Delta m\neq0$, in gray scale. Note that
due to the effect of the $\Delta m$ values between 5.0 and 5.5 are possible for \hv, while the same stands for $\beta\in(0.041,0.706)$ mag per degree.
The continuous lines (red in electronic version) show the area that contain 68.3, 95.5, and 99.7 \% of the solutions.  
              }
         \label{example_lc}
   \end{figure}

In other words, for objects with $\Delta m>0$ we have 10,000 different solutions for Eq. \ref{linear}. We computed the average of the solutions
for \hv~and $\beta$ and assumed these values as the most likely result. A graphical representation of the procedure is shown in Fig. \ref{example_lc}.
In the figure the left panel shows the representative phase curve along with the data points and their errors, while the right panel shows a two-dimensional
histogram showing the phase-space covered by the 10,000 solutions.
This method allowed us to explore the solution space finding some interesting results, such as those unexpected cases with $\beta<0$, {\bf which
we will discuss in Sect. \ref{Disc}.}

All results are shown in Table \ref{table:1}. The table reports observed object, \hv~and $\beta$, the number of points used in the fits, the
light-curve amplitude, and the references to the works whose reported magnitudes were used. 

\section{Results}

\begin{figure}
\centering
\includegraphics[width=\hsize]{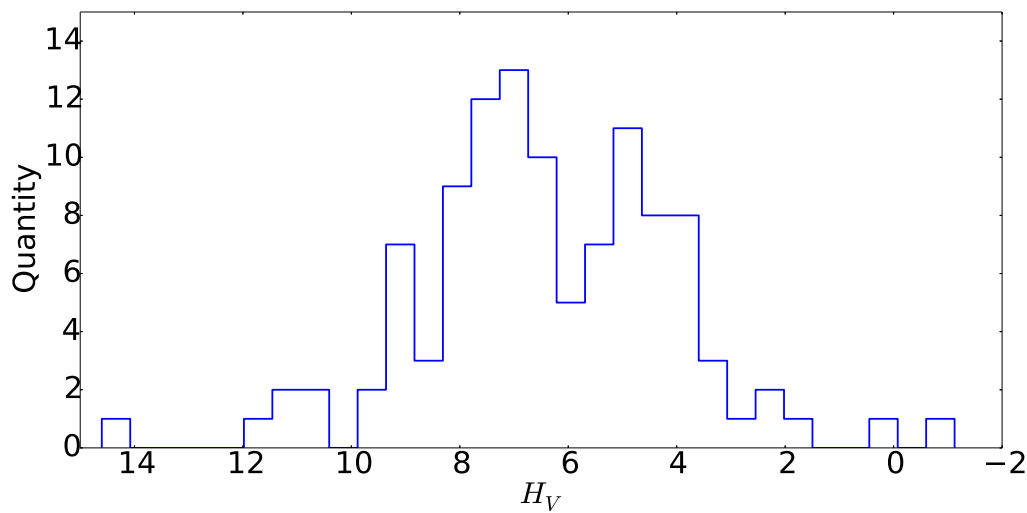}
\caption{Histogram showing the distribution of obtained \hv.}
\label{hmag_histo}
\end{figure}

\begin{figure}
\centering
\includegraphics[width=\hsize]{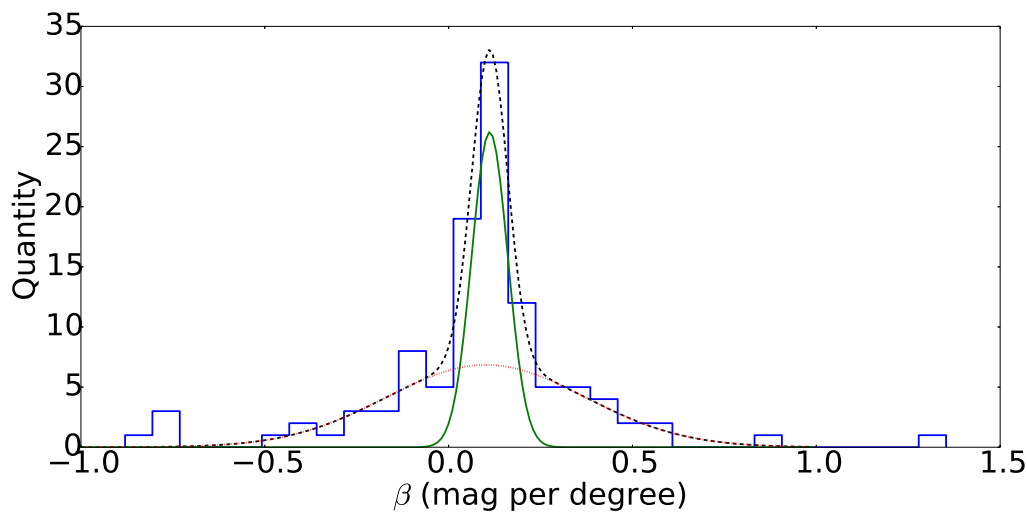}
\caption{Histogram showing the distribution of obtained $\beta$.
{\bf The dashed black line is the better fit to the distribution, modeled as the sum of two Gaussian distributions (see text). Each individual Gaussian distribution is 
shown in continuous green and dotted red lines.}}
\label{beta_histo}
\end{figure}

We measured \hv~and $\beta$ for a total of 110 objects. Figure \ref{hmag_histo} shows the distribution of \hv~resulting from applying our procedure.
{\bf The distribution looks bi-modal}, with one peak, the larger one, at \hv~$\approx7$ and a second one at \hv~$\approx5$.
Our results cover a range from a minimum of \hv~$=14.6$ (2005 UJ$_{438}$) up to a maximum of -1.12 for Eris. The average value is 6.39, while the median is 6.58. 
The distribution of $\beta$ is shown in Fig. \ref{beta_histo}. The average value is 0.09 mag per degree, while the median is 0.10 mag per degree, 
with a minimum of -0.88 mag per degree for 2003 GH$_{55}$ and a maximum of 1.35 mag per degree for 2004 GV$_{9}$. Almost
60 \% of the values fall within 0.01 and 0.23 mag per degree.

{\bf 
Curiously, the distribution shown in Fig. \ref{beta_histo} seems to be the combination of two different distributions, one wide and shallow and a second one sharp and tall.
To test this possibility we assumed that the distribution could be fitted by a sum of two Gaussian distributions 
$$F(\beta)=C_1e^{-\frac{(\beta-\beta_1)^2}{2\sigma_1^2}}+C_2e^{-\frac{(\beta-\beta_2)^2}{2\sigma_2^2}},$$ where $C_i$, $\beta_i$, and $\sigma_i$ are free parameters.

We run a minimization script from python ({\tt scipy.optimize.leastsq}) obtaining all six free parameters: $C_1=6.8$, $\beta_1=0.27$ mag per degree, $\sigma_1=0.10$ mag per degree, 
and $C_2=26.2$, $\beta_1=0.05$ mag per degree, $\sigma_1=0.11$ mag per degree.
The best fitting $F(\beta)$ is shown in Fig. \ref{beta_histo}, along with the two components. The two-Gaussian model describe rather well the distribution of $\beta$, both with 
similar modes but different widths. In the Discussion we will come back to this model.
}

Next, we compare our results with those of a few selected works: 
Rabinowitz et al. (2007), Perna et al. (2013), and B\"ohnhardt et al. (\cite{boenh14}); and then 
we search for correlations among our results (\hv, colors, $\beta$), orbital elements (semi-major axis,
eccentricity, inclination), the absolute magnitudes used in the ``TNOs are cool'' Herschel Space Observatory key project and their
measured geometric albedos, and the light-curve amplitude $\Delta m$.
Orbital elements for each object were obtained from the Lowell Observatory\footnote{\tt ftp://ftp.lowell.edu/pub/elgb/astorb.html}.

\subsection{Comparison with selected works} 
On one hand we selected Rabinowitz et al. (2007, Ra07) because it has the most dense phase curves reported for 25 outer 
solar system objects, while on the other hand Perna et al. (2013, Pe13) and B\"ohnhardt et al. (\cite{boenh14}, Bo14) present results in support for the HSO
``TNOs are cool'' key project. The three works analyse their data following different criteria: Ra07 observed each target on many occasions,
even attempting to obtain rotational properties. If a rotational light-curve could be determined, the data were corrected removing the short term variability,
the remaining data were then rebinned in $\alpha$, and then the phase curves were constructed. Pe13, using less dense data, computed phase curves for a few objects while for
these objects with no enough data average values of $\beta$ were assumed. Bo14 only used average values of $\beta$.

We report in Table \ref{table:comp} the values of \hv~along with the estimated $\beta$, if not used an average value. Note that our phase curves include
the data reported in these three works.

Overall, the agreement is very good among the four works. Nevertheless, there are some values that differ beyond three sigma. For clarity we report 
here these differences {\bf (shown in boldface in  Table \ref{table:comp})}. With Ra07 Makemake (\hv~and $\beta$) and Sedna (\hv); with Pe13 2005 UJ$_{438}$ (\hv) 
and Varda (\hv); with Bo14 2003 GH$_{55}$, 2004 PG$_{115}$, and Okyrhoe. In this last case the differences are only in 
\hv~because these authors did not compute the phase curve, but used instead average values of $\beta$ to obtain absolute magnitudes.
Also, the errors in our data are somewhat larger than those in Ra07, Pe13 and Bo14. We will come back to this issue in the discussion.

\subsection{Correlations}
We searched for possible correlations among pairs of variables. We define here a variable as any given set of quantities representing the population, for instance
the variable $\beta$ is the set of phase coefficients of the TNOs sample. 
The correlations were explored using a Spearman test, which has the vantage of being non-parametric relying on ordering the data according to rank and 
running a linear regression through those ranks. The test returns two values $r_s$, which 
gives the level of correlation of the tested variables, $|r_s|\approx1$ indicates correlated quantities, while $|r_s|\rightarrow0$ indicates uncorrelated data.
The reliability of $r_s$ is given by $P_{r_s}$ which indicates the probability of two variables being uncorrelated, in practical terms, the closer
$P_{r_s}$ is to zero, the more likely the result provided by $r_s$ becomes.

One disadvantage of the Spearman test is that it does not consider the errors in the variables. To overcome this issue we proceeded
as follows: let us assume we are trying to find the correlation among a set $\{x_j,y_j\}$, where each quantity $x_j$ ($y_j$) has an error of $\sigma_{x_j}$
($\sigma_{y_j}$), $j$ running from 1 up to $N$. Now we create 10,000 correlations by creating new sets $\{{x_j}_i,{y_j}_i\}$, where 
${x_j}_i=x_j+rand_i\times\sigma_{x_j}$, likewise for $y_j$. {\bf In this case $rand_i$ is a random number drawn from a normal distribution in $[-1,1]$.} 
The random number in $x_j$ is not necessarily the same as in $y_j$.

\begin{figure}
   \centering
   \includegraphics[angle=0,width=\hsize]{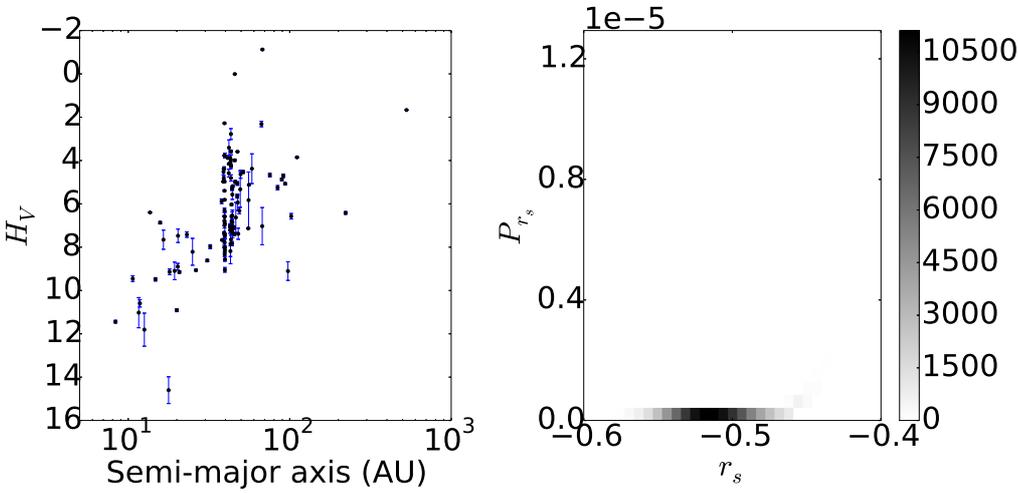}
      \caption{Left: scatter plot of \hv~vs. semi-major axis. Right: Outcome of the 10,000 realizations in form of a two-dimensional histogram in $r_s$ and 
$P_{r_s}$ which shows the phase space where the solutions lie. In a few cases it is relatively clear that a correlation might exists, while in some other cases 
large excursions are seen, which indicate that an false correlation could arise in the case of large errors.}
         \label{corr8}
   \end{figure}

\begin{figure}
   \centering
   \includegraphics[angle=0,width=\hsize]{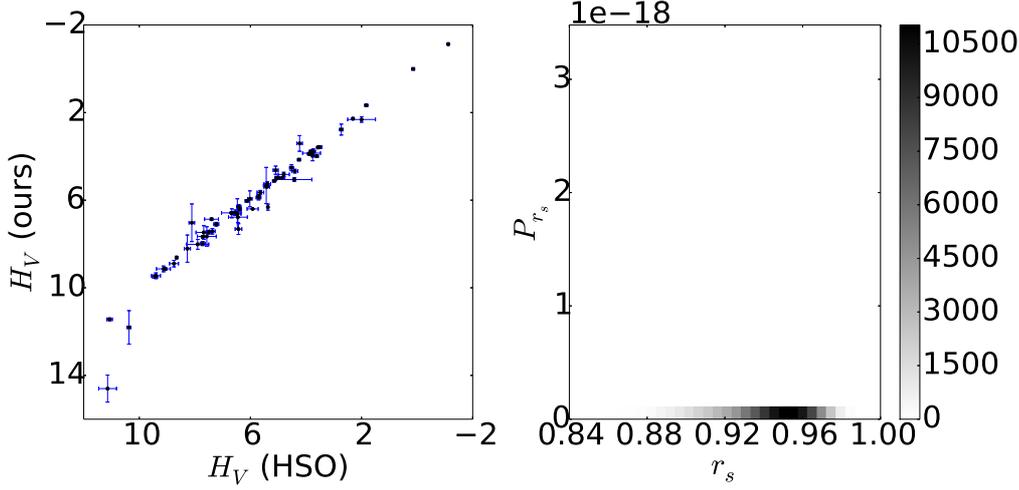}
      \caption{Left: scatter plot of \hv~as measured by us vs (ours). \hv~as used within the ``TNO's are cool'' program (HSO). 
Right: Two-dimensional histogram showing the most likely correlations.}
         \label{corr11}
   \end{figure}

\begin{figure}
   \centering
   \includegraphics[angle=0,width=\hsize]{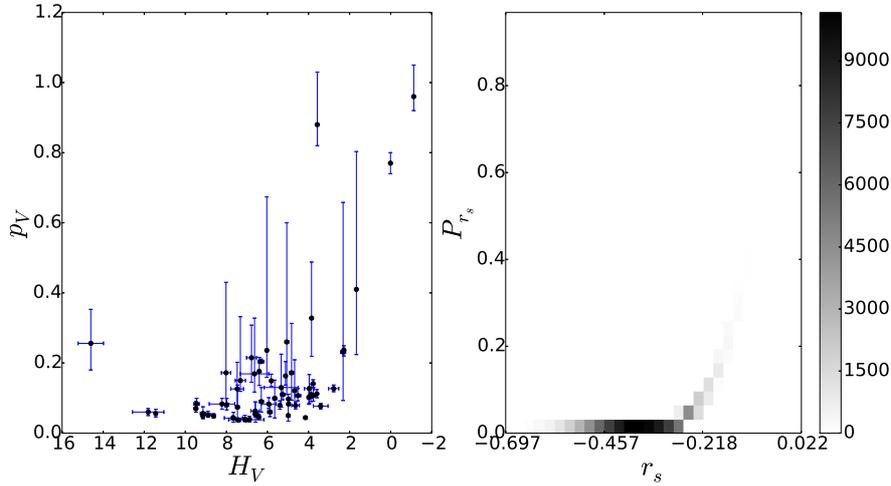}
      \caption{Left: scatter plot of \hv~vs. the geometric albedo measured by the ``TNOs are cool'' program. 
Right: Two-dimensional histogram showing the most likely correlations.}
         \label{corr13}
   \end{figure}

\begin{figure}
   \centering
   \includegraphics[angle=0,width=\hsize]{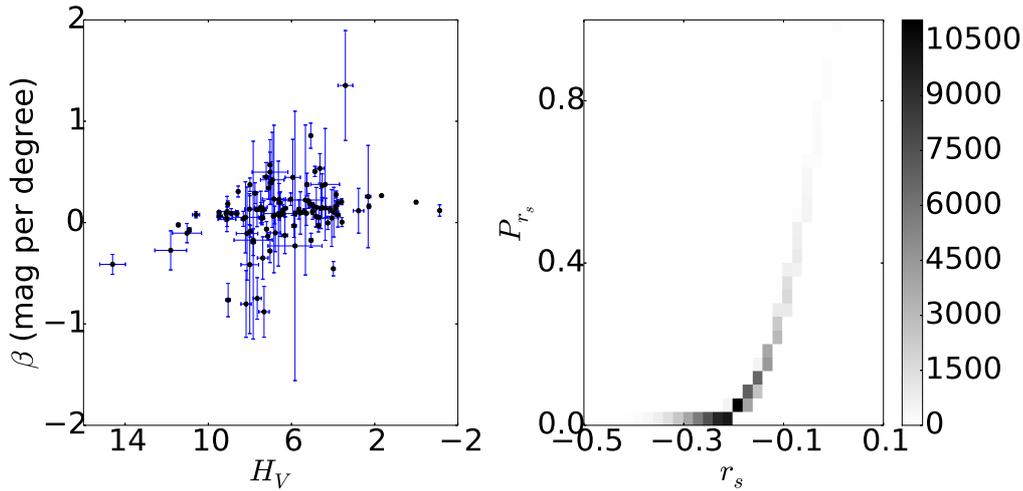}
      \caption{Left: scatter plot of $\beta$ vs. \hv. Right: Two-dimensional histogram showing the most likely correlations.}
         \label{corr1}
   \end{figure}

\begin{figure}
   \centering
   \includegraphics[angle=0,width=\hsize]{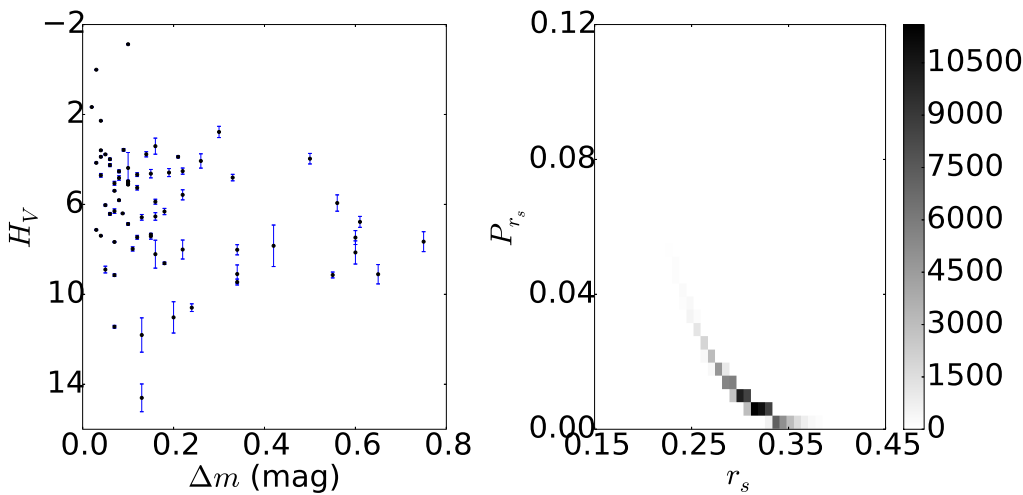}
      \caption{Left: scatter plot of \hv~vs. $\Delta m$. Right: Two-dimensional histogram showing the most likely correlations.}
         \label{corr12}
   \end{figure}

\begin{figure}
   \centering
   \includegraphics[angle=0,width=\hsize]{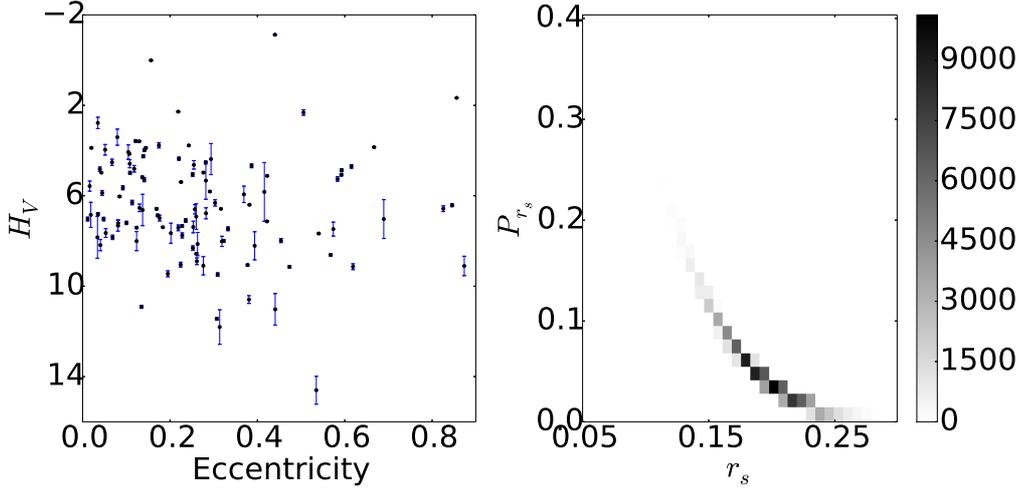}
      \caption{Left: scatter plot of \hv~vs. eccentricity. Right: Two-dimensional histogram showing the most likely correlations.
              }
         \label{corr9}
   \end{figure}

\begin{figure}
   \centering
   \includegraphics[angle=0,width=\hsize]{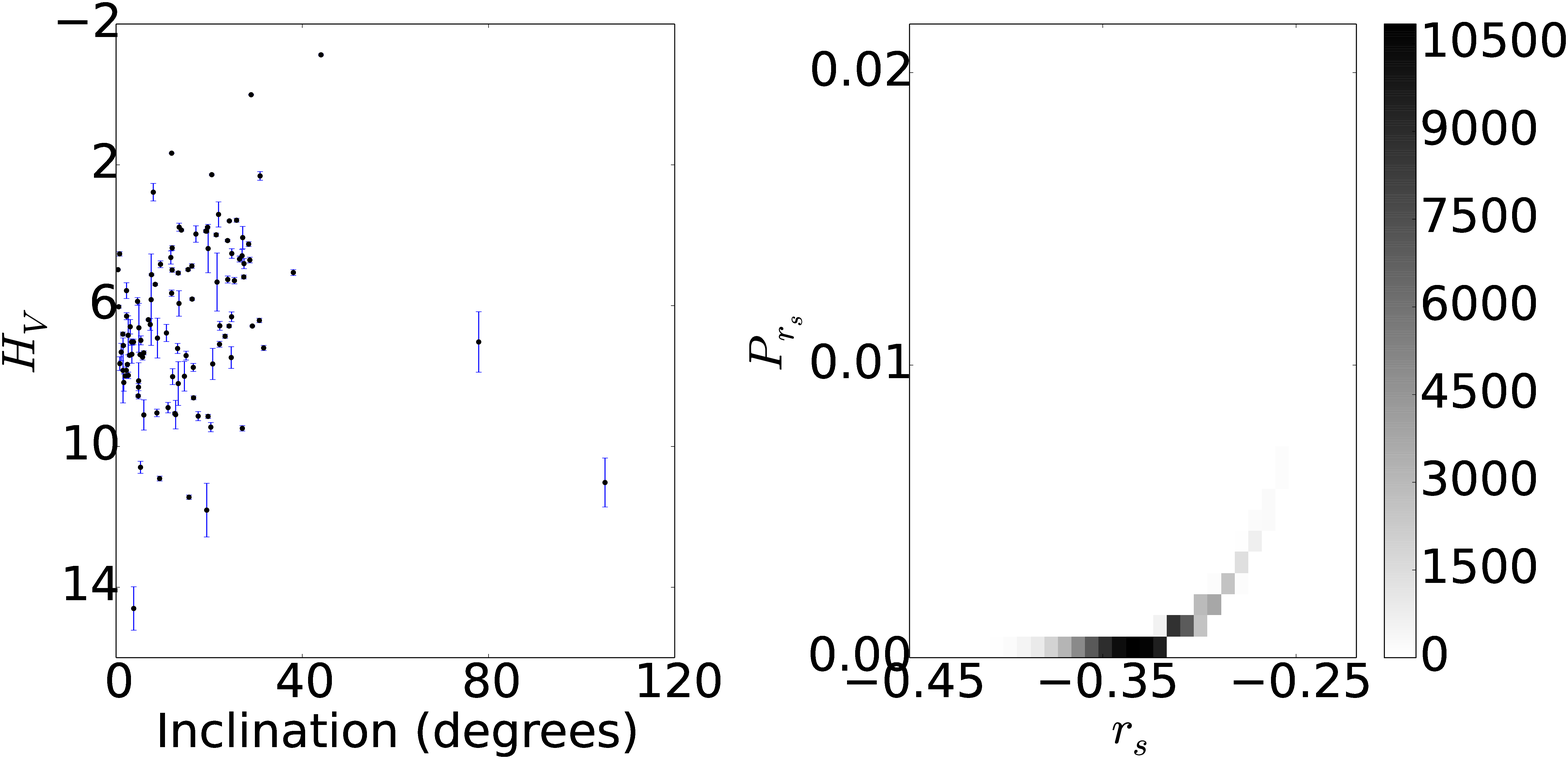}
      \caption{Left: scatter plot of \hv~vs. inclination. Right: Two-dimensional histogram showing the most likely correlations.
              }
         \label{corr10}
   \end{figure}

After performing the 10,000 correlations we finish with a set $\{{r_s}_i,{P_{r_s}}_i\}$, which are displayed in form of density plots to show
the likelihood of the correlation to held against the error bars. All relevant results are displayed in Figs. \ref{corr8} - \ref{corr10}.
Table \ref{table:corr} shows the result of the correlation tests: the first column shows the 
variables tested, second and third column show the nominal values of $r_s$ and $P_{r_s}$ {\bf (those where the errors were not accounted for)}, while the last
column reports our interpretation of the density plots of whether the correlation exists or not.
\begin{table}
\caption{Correlations}\label{table:corr}
\centering
\begin{tabular}{rccl}
\hline \hline
Variables       & $r_s$ & $P_{r_s}$ & Correlation \\
\hline
\hv~vs. $a$		& -0.517 & $7.6\times10^{-9 }$  & yes *  \\
\hv(ours) vs. \hv(HSO)	&  0.987 & $7.1\times10^{-51}$	& yes  \\
\hv~vs. $p_{\rm V}$	& -0.509 & $1.8\times10^{-5 }$	& yes  \\
$\beta$ vs. \hv 	& -0.379 & $4.5\times10^{-5 }$	& weak \\
\hv~vs. $\Delta m$      &  0.359 & 0.0020               & weak \\

\hv~vs.  Inclination	& -0.335 & 0.0003               & weak \\
\hv~vs. $e$		&  0.207 & 0.0299               &  no  * \\

$\beta$ vs. $\Delta m$ 	& -0.141 & 0.2358               &  no  \\
$\beta$ vs. $p_{\rm V}$	&  0.011 & 0.9341               &  no  \\
\hv~vs. V-R     	&  0.185 & 0.0532		&  no  \\

$\beta$ vs. V-R 	&  0.090 & 0.3474 		&  no  \\
$\beta$ vs. $a$ 	&  0.233 & 0.0142               &  no  \\
$\beta$ vs. $e$ 	&  0.137 & 0.1525               &  no  \\
$\beta$ vs. Inclination	&  0.140 & 0.1450               &  no  \\
\hline                                                         
\end{tabular}
\tablefoot{* Observational bias}
\end{table}

For the scope of the present work we decided not to separate our sample into the sub-populations that appear among Centaurs and TNOs because 
dividing a sample of 110 objects into smaller samples will just {\bf decrease the 
statistical significance of any possible result.} Furthermore, should any real difference arise among any subgroup, this would clearly be seen in any of the tests proposed here.
{\bf Such as the fact that no large Centaurs are known, or that the so-called Classic TNOs have low inclinations and then to be smaller in size than other subpopulations
of TNOs.}
Below we report the most interesting findings of the search for correlations. Thereafter, we discuss on some individual cases that showed interesting 
or anomalous behaviours.

\smallskip
\noindent
{\bf \hv~vs. semi-major axis:}
Figure \ref{corr8} shows the correlation between absolute magnitude and semi-major axis. This correlation is due to observational bias and accounts for
the lack of faint objects detected at large heliocentric distances, while no bright Centaur (defining loosely a Centaur as an object with semi-major axis below 
30 AU) is known to exist.

\smallskip
\noindent
{\bf \hv(ours) vs. \hv(HSO):}
In this case we compared our computed magnitudes with those used by the Herschel Space Observatory ``TNOs are cool'' key project. The correlation is 
close to 1 (Fig. \ref{corr11}), although it is possible to see a small departure at the faint end with two objects with significantly smaller \hv, 
they are (250112) 2002 KY$_{14}$ (\hv~$=11.808\pm0.763$) and (145486) 2005 UJ$_{438}$ (\hv~$=14.602\pm0.617$). 
In the first case we revised the data without finding any evident problem and we trust the value to be correct, while in the second case some care 
should be taken because the minimum value of phase angle is about 5.8\degs~leaving most of the phase curve under sampled, which might be affecting 
the value of $\beta$.

{\bf For the sake of comparison we fitted a linear function to the data according to \hv(ours)$=a+b\times$\hv(HSO), obtaining $b= 1.06\pm0.03$ and $a=-0.27\pm0.17$.
This indicates that, although \hv(ours) are very similar to \hv(HSO) they are not identical. This difference between our \hv~and those used by the ``TNOs are cool'' team
are probably due to the fact that some of theirs were computed using single observations and assuming an average $\beta$.}

\smallskip
\noindent
{\bf \hv~vs. $p_{\rm V}$:}
Figure \ref{corr13} shows that there exists a correlation between the absolute magnitude and the geometric albedo, the brigher the object the largest 
the albedo. This is probably reflecting the fact that brighter objects tend to be the larger ones in size as well, and therefore are able to retain part of the
original volatiles more reflective species that smaller objects cannot.

\smallskip
\noindent

\smallskip
\noindent
{\bf $\beta$ vs. \hv:}
\hv~seems to have a weak anti-correlation with $\beta$ indicating that brighter objects have larger positive slopes
than fainter ones. From Fig. \ref{corr1} one interesting details arise: there are a few objects with $\beta<0$ (see also Fig. \ref{beta_histo}), 
even considering errorbars and light-curve amplitude (see Table \ref{table:1}). This issue deserve further study and observations.
From the density map the weak correlation seems quite consistent within the errors in \hv~and $\beta$.

\smallskip
\noindent
{\bf \hv~vs. $\Delta m$:}
There is a weak correlation between absolute magnitude and $\Delta m$ pointing that brighter objects 
tend to have lower $\Delta m$. Interestingly, among the faint object (fainter than \hv~$=10$) no large ($>0.25$) amplitudes are found
(Fig. \ref{corr12}). It should be remembered that, although faint objects, these are usually in the range 50 to 100 km (see 
{\tt http://public-tnosarecool.lesia.obspm.fr/}).

{\bf In a previous work, Duffard et al. (\cite{duffa09}) presented a similar value for this correlation. Also using their results (their Fig. 6) it is possible to see that objects
with densities lower than 0.7 g cm$^{-3}$ are unlikely in hydrostatic equilibrium and therefore could have large $\Delta m$, which is not reflected in our
Fig. \ref{corr12}. These density correspond to $\approx400$ km (from Fig. S7 in Ortiz et al. \cite{ortiz12}) which is roughly \hv~$\approx5.4$. Brighter, possibly larger,
objects are in hydrostatic equilibrium and their shapes are better described by Mclaurin spheroids whose $\Delta m$ are harder to measure due to their symmetry around 
the minor axis.}

\smallskip
{\bf \hv~vs. Eccentricity and Inclination}
There are two curious cases (Fig. \ref{corr9} and \ref{corr10}). The first one, \hv~vs. eccentricity indicates that fainter objects tend to have higher 
eccentricities. This is an observational bias because faint objects are more easily observed close to perihelion, favoring objects with high eccentricties. 
The second one, \hv~vs. inclination, shows also a weak tendency of fainter objects having smaller inclinations. This might be reflecting the known fact
that among the so-called ``Classical'' trans-Neptunian belt are found two subpopulations, the hot and cold (from dynamical considerations) where the cold, 
low-inclination population does not have objects as large as the hot, high-inclination, population.
Note that although both tendencies seem significant over the 2-sigma level ($>95.5$ \%), only one seems closer to be a correlation having $|r_s|>0.3$. 

\paragraph{\bf Other results:}
None of the other pairs of variables explored show any significant correlation, therefore their plots are not reported. 






\smallskip
\paragraph{\bf Interesting objects}
In this paragraph we describe some objects that deserve more discussion.

\noindent
{\em 2060 Chiron,} Meech and Belton (1989) detected a coma surrounding Chiron, this result probably influenced the interpretation of
latter stellar occultations results (e.g. Bus et al. 1996) that detected secondary events which were associated to jets of material ejected from
the surface. A recent re-analysis of all stellar occultation data, along with new photometric data, suggests that Chiron possesses
a ring system (Ortiz et al. 2015). {\bf Both phenomena, cometary-like activity and the possible ring system, affect the photometric data obtained from Chiron, including 
the way the photometric measurements are performed, thus increasing the scattering in the phase curve}.

\noindent
{\em 10199 Chariklo,} Braga-Ribas et al. (2014) detected a ring system around Chariklo using data from a stellar occulation. This result
helped to interpret long-term changes in photometric and spectroscopic data (Duffard et al. 2014) such as the secular variation
in reduced magnitude (Belskaya et al. 2010) and the disappearance of a water-ice absorption feature in its near-infrared spectrum 
(Guilbert et al. 2009). As with the case of Chiron, the phase curve of Chariklo does not follow a linear trend.

\noindent
{\em Bright objects,} those with \hv~brighter than 3 have $\beta$ between 0.11 and 0.27 mag per degree. 
Spectroscopically it is known that these objects (2007 OR$_{10}$, Eris, Makemake, Orcus,
Quaoar, and Sedna) are very different, Eris and Makemake display CH4 absorption features while Orcus, Quaoar, and 2007 OR$_{10}$ show water ice and probably 
some hydrocarbons. Therefore, particle size or compaction could play a more important role than composition on the phase curves.

\section{Discussion and Conclusions}\label{Disc}

We have observed 56 objects, six of them with no reported magnitudes in the literature, to the best of our knowledge. We combined
these new $V$ and $R$ magnitudes with an extensive bibliographic survey to compute absolute magnitudes and phase coefficients. 
In total we report \hv~and $\beta$ for 110 objects.
Some of these objects had phase curves already reported, nevertheless it is important to include new data, always keeping in mind that we are combining
data from different apparitions for the same object and that surface conditions might have changed between observations.

Regarding the distribution of $\beta$, Fig. \ref{beta_histo} clearly shows a quasi symmetric distribution. {\bf The maximum and mode, coincides, to the second decimal place, 
with the average and median values: 0.10 mag per degree.
We tested the hypothesis of having a two-population distribution, assuming that each population could be described by a Gaussian function. The fit to the data is quite good,
but does this indicate the existence of two {\it real} subpopulations? One possible explanation regards the quality of the data: A ``high-quality'' sub-sample clusters with a mode of
$\beta_2=0.11$ mag per degree within a sharp distribution, while ``low-quality'' data is more spread, but with a very similar mode ($\beta_2=0.10$ mag per degree), considering as
high-quality data these with small errors, precise $\beta$, and with (at least) an estimation of $\Delta m$. Unfortunately, this is not strictly the case, as some of these objects
fall within the wide and shallow distribution, far from the sharp and tall distribution. Therefore, even if very tempting, we cannot use the sharp distribution as representative 
of the whole population as we could introduce undesired biases in the results. Moreover, most of the objects fall within the wide distribution, 59 \%, while 41 \% fall within the sharp one.}

It is clear that there is not one representative value of $\beta$ for the whole population. Therefore,
the use of average values of $\beta$ to compute \hv~should be regarded with caution.
The phase coefficients range from -0.88 up to 1.35 mag per degree. 
On the extreme positive side, the two objects (1996~GQ$_{21}$ and 2004~GV$_{9}$)
have large errors associated. Among the extreme
negative values, there are six objects (1998~KG$_{62}$, 1998~UR$_{43}$, 2002~GP$_{32}$, 2003~GH$_{55}$, 
2005~UJ$_{438}$, and Varda which have $\beta<0$ even considering three times the error. 
Most of these cases are objects whose data are sparse and with few points. Two of them, UJ438 and Varda, have estimated light-curve amplitude while the rest 
has no reported value to the best of our knowledge. 

{\bf We are no aware of any physical mechanism that could explain a $\beta<0$ using scattering models. 
There are some components of the light that could be negative, such as the incoherent second scattering order (Fig. 21 in Shkuratov et al. \cite{shkur02})
which is nonetheless non dominant, especially for the low values of $\alpha$ we can observe TNOs.}

These extremes values, either positive or negative, could be due to yet undetected phenomena 
such as rotational modulation poorly determined, ring systems, or cometary-like activity and deserve more observations. 

There are some phase curves that clearly do not follow a linear trend. Those of Chiron and Chariklo, in fact, do not follow any particular trend at all.
It is convenient to bear in mind that the photometric models to understand the photometric behaviour of phase curves were thought for
objects with nothing else than their bare surface to reflect/scatter/absorb photons. In the case of these possibly ringed systems the
reflected light detected on Earth depends not only on the scattering properties of the material covering Chiron, or Chariklo, but as well
of the particles in the rings and the geometry of the system. With this in mind, we propose that one criterion to seek candidates to bear
ring systems is to search for these ``non-linear'' behaviour of the phase curve. As examples, based on the dispersion seen in their phase curves
we propose that 1996 RQ$_{20}$, 1998 SN$_{165}$, or  2004 UX$_{10}$ as candidates for further studies,
among other objects.

The correlations were discussed in their respective paragraphs. Overall, some of them are associated to observational biases (\hv~and semi-major axis; 
\hv~and eccentricity), other can be interpreted in terms of known properties of the TNO region (\hv~and inclination) while the rest can be considered as weak or
non-existing and deserving more data, especially going deeper into the faint end of the population. We do not confirm a proposed anti-correlation 
between albedo and phase coefficient (see Belskaya et al. 2008 and references therein). One special note is deserved by the anti-correlation found between \hv~and
$p_{\rm V}$, it would seem that the correlation is driven principally by the brighter objects. We ran the same test discarding those objects brighter than
3, and those associated to the Haumea dynamical group because they form a group that stands aside with particular surface properties,
and the relation still holds, $r_s=-0.356,~P_{r_s}=0.0092$. Although the correlation does become weaker, without reaching a $3-\sigma$ level, there seems 
to exists a trend of brighter objects to have larger geometric albedos. An in-depth physical explanation remains yet to be formulated.

Finally, the errors reported in \hv~are in some cases larger than previous works. This is reflecting the heterogeneity of the sample, how the 
effect of the rotational variability is considered, and the weighing of the data while performing the linear fits. Note, for instance, that
all of the objects with $\sigma_{H_{\rm V}}>0.1$ mag have either less than 10 data points or $\Delta m>0.1$ mag. 
Taking this into consideration our results are more accurate, although not as precise, than previous works and probably more realistic, with the exception of the strategy followed by
Rabinowitz et al. (\cite{rabin07}).
This work represents the first release of data taken at seven different telescopes in six observatories between late 2011 and mid 2015 which represents a large effort. 
It is important to mention that more observations are ongoing.

\begin{acknowledgements}
AAC acknowledges support via diverse grants to FAPERJ and CNPq. 
JLO acknowledges support from the spanish Mineco grant AYA-2011-30106-CO2-O1, 
from FEDER funds and from the Proyecto de Excelencia de la Junta de Andalucía, J.A. 2012-FQM1776.
R.D. acknowledges the support of MINECO for his Ram\'on y Cajal Contract.
The authors would like to thank Y. Jim\'enez-Teja for technical support and P.H. Hasselmann for helpful discussions regarding phase curves.
We are grateful to O. Hainaut whose comments helped to improve the quality of this manuscript.
\end{acknowledgements}


\Online

\begin{appendix} 
\section{Tables}

\begin{longtab}
\begin{longtable}{r c c c c c c c l}
\caption{\label{table:obs}Observations}\\             
\hline\hline
  Object    &     V    &     R     &   Night   &  $r$ (AU) &  $\Delta$ (AU) & $\alpha$ (degress)&  Telescope &  Notes \\ 
\hline
\endfirsthead
\caption{continued.}\\
\hline\hline
  Object    &     V    &     R     &   Night   &  $r$ (AU) &  $\Delta$ (AU) & $\alpha$ (degress)&  Telescope &  Notes \\ 
\hline
\endhead
\hline
\endfoot
  24835      1995 SM$_{55 }$  & 19.898$\pm$0.216  &   19.170$\pm$0.132  &  2012-12-09  & 38.4165  &  37.6015 & 0.8285 & CAHA2.2    &(1)    \\
  26181      1996 GQ$_{21 }$  & 21.536$\pm$0.192  &   20.900$\pm$0.155  &  2014-05-29  & 42.6600  &  41.6917 & 0.4020 & SOAR       &(1)    \\
  26181      1996 GQ$_{21 }$  & 21.760$\pm$0.217  &   20.516$\pm$0.086  &  2013-06-10  & 42.3455  &  41.4599 & 0.6737 &  INT       &       \\
  26181      1996 GQ$_{21 }$  & 21.775$\pm$0.233  &                     &  2013-06-11  & 42.3464  &  41.4692 & 0.6939 &  INT       &(1)    \\
  40314      1999 KR$_{16 }$  & 21.871$\pm$0.132  &   20.767$\pm$0.083  &  2013-06-03  & 35.3552  &  34.4522 & 0.7536 & CAHA3.5    &       \\
  40314      1999 KR$_{16 }$  & 21.586$\pm$0.091  &   20.905$\pm$0.085  &  2014-04-02  & 35.2260  &  34.4426 & 1.0274 & SOAR       &(1)    \\
  47171      1999 TC$_{36 }$  & 20.373$\pm$0.134  &   19.504$\pm$0.087  &  2013-09-03  & 30.5720  &  29.8969 & 1.4249 &  OSN       &(1)    \\
  47932      2000 GN$_{171}$  & 21.313$\pm$0.070  &   20.852$\pm$0.069  &  2014-04-02  & 28.4086  &  27.6404 & 1.3140 & SOAR       &(1)    \\
  82075      2000 YW$_{134}$  & 21.219$\pm$0.401  &   21.039$\pm$0.407  &  2012-12-09  & 44.6975  &  44.1306 & 1.0432 & CAHA2.2    &(1)    \\
  82158      2001 FP$_{185}$  & 21.407$\pm$0.489  &   20.779$\pm$0.297  &  2013-04-14  & 35.4526  &  34.4818 & 0.4155 &  OSN       &       \\
  82158      2001 FP$_{185}$  & 22.354$\pm$0.631  &   20.723$\pm$0.399  &  2013-05-11  & 35.4714  &  34.5972 & 0.8229 & CAHA3.5    &(1)    \\
             2001 KD$_{77 }$  & 21.799$\pm$0.181  &   21.121$\pm$0.110  &  2013-06-03  & 35.9812  &  35.0132 & 0.4968 & CAHA3.5    &       \\
 139775      2001 QG$_{298}$  & 22.068$\pm$0.239  &   22.076$\pm$0.202  &  2013-07-17  & 31.7844  &  31.6575 & 1.8231 & CAHA3.5    &(1)    \\
  55565      2002 AW$_{197}$  & 20.720$\pm$0.233  &   19.849$\pm$0.116  &  2013-04-15  & 46.0579  &  45.5946 & 1.1096 &  OSN       &       \\
  55565      2002 AW$_{197}$  &                   &   19.900$\pm$0.173  &  2013-04-17  & 46.0589  &  45.6247 & 1.1283 &  OSN       &(1,2)  \\
             2002 GH$_{32 }$  & 21.988$\pm$0.203  &   21.726$\pm$0.303  &  2013-06-11  & 43.4742  &  42.6321 & 0.7500 &  INT       &(1)    \\
             2002 GP$_{32 }$  & 22.124$\pm$0.087  &   21.788$\pm$0.069  &  2013-06-11  & 32.3989  &  31.4001 & 0.3222 &  INT       &(1)    \\
             2002 GP$_{32 }$  & 21.824$\pm$0.631  &   22.023$\pm$0.037  &  2013-07-16  & 32.4100  &  31.6764 & 1.2527 & CAHA3.5    &       \\
  95626      2002 GZ$_{32 }$  & 20.133$\pm$0.201  &   19.829$\pm$0.157  &  2013-04-14  & 18.5160  &  17.6439 & 1.5778 &  OSN       &       \\
 119951      2002 KX$_{14 }$  &                   &   20.375$\pm$0.253  &  2013-06-10  & 39.2847  &  38.2818 & 0.2282 &  INT       &       \\
 250112      2002 KY$_{14 }$  & 19.943$\pm$0.136  &   19.383$\pm$0.090  &  2012-12-08  & 9.5689   &  8.8098  & 3.9111 & CAHA2.2    &       \\
 250112      2002 KY$_{14 }$  & 20.413$\pm$0.091  &   19.550$\pm$0.100  &  2012-12-11  & 9.5725   &  8.8448  & 4.1247 & CAHA2.2    &       \\
 307261      2002 MS$_{4  }$  & 20.064$\pm$0.053  &   18.907$\pm$0.073  &  2013-06-03  & 47.0005  &  46.0946 & 0.5670 & CAHA3.5    &(3)    \\
 307261      2002 MS$_{4  }$  & 20.184$\pm$0.270  &   20.406$\pm$0.616  &  2013-05-10  & 47.0046  &  46.3190 & 0.9151 & CAHA3.5    &(3)    \\
  55637      2002 UX$_{25 }$  & 19.474$\pm$0.106  &   19.632$\pm$0.116  &  2011-10-31  & 41.4407  &  40.4513 & 0.1105 & CAHA2.2    &(1)    \\
  55637      2002 UX$_{25 }$  & 20.203$\pm$0.072  &   19.606$\pm$0.044  &  2012-10-16  & 41.3080  &  40.3379 & 0.3268 & CAHA2.2    &       \\
  55637      2002 UX$_{25 }$  & 20.286$\pm$0.084  &   20.164$\pm$0.097  &  2012-12-11  & 41.2868  &  40.5805 & 0.9553 & CAHA2.2    &       \\
  55637      2002 UX$_{25 }$  & 19.800$\pm$0.085  &   19.545$\pm$0.077  &  2013-09-02  & 41.1853  &  40.6445 & 1.1965 &  OSN       &(1)    \\
  55638      2002 VE$_{95 }$  & 20.490$\pm$0.104  &   19.713$\pm$0.106  &  2012-12-11  & 28.8622  &  27.9090 & 0.4964 & CAHA2.2    &       \\
  55638      2002 VE$_{95 }$  & 21.223$\pm$0.416  &   20.003$\pm$0.171  &  2011-10-31  & 28.6992  &  27.9126 & 1.2313 & CAHA2.2    &(1)    \\
 119979      2002 WC$_{19 }$  & 21.099$\pm$0.402  &   21.281$\pm$0.563  &  2012-12-09  & 41.7521  &  40.7709 & 0.1179 & CAHA2.2    &(1,3)  \\
 127546      2002 XU$_{93 }$  & 21.716$\pm$0.401  &   21.131$\pm$0.294  &  2012-12-11  & 21.5317  &  20.9261 & 2.0983 & CAHA2.2    &       \\
 127546      2002 XU$_{93 }$  & 21.565$\pm$0.334  &   21.868$\pm$0.535  &  2012-12-10  & 21.5310  &  20.9301 & 2.1084 & CAHA2.2    &(1)    \\
 127546      2002 XU$_{93 }$  & 21.214$\pm$0.260  &   21.033$\pm$0.213  &  2012-12-08  & 21.5296  &  20.9391 & 2.1302 & CAHA2.2    &       \\
 127546      2002 XU$_{93 }$  & 21.007$\pm$0.144  &   21.180$\pm$0.162  &  2012-10-16  & 21.4937  &  21.3872 & 2.6493 & CAHA2.2    &       \\
 120132      2003 FY$_{128}$  & 20.034$\pm$0.386  &   19.541$\pm$0.299  &  2013-04-15  & 39.1861  &  38.1980 & 0.2555 &  OSN       &       \\
 120132      2003 FY$_{128}$  & 21.063$\pm$0.200  &   20.254$\pm$0.186  &  2013-04-16  & 39.1865  &  38.2004 & 0.2731 &  OSN       &       \\
 120178      2003 OP$_{32 }$  & 20.269$\pm$0.124  &   19.794$\pm$0.166  &  2012-09-16  & 41.7560  &  40.8552 & 0.6137 & CAHA2.2    &       \\
 120178      2003 OP$_{32 }$  & 20.084$\pm$0.139  &   20.248$\pm$0.185  &  2012-09-17  & 41.7562  &  40.8621 & 0.6310 & CAHA2.2    &(1)    \\
 120178      2003 OP$_{32 }$  & 20.168$\pm$0.100  &   19.950$\pm$0.082  &  2011-09-24  & 41.6606  &  40.8276 & 0.7714 & CAHA2.2    &       \\   
 120178      2003 OP$_{32 }$  & 20.044$\pm$0.155  &   19.891$\pm$0.215  &  2011-09-25  & 41.6609  &  40.8367 & 0.7886 & CAHA2.2    &       \\
 120178      2003 OP$_{32 }$  & 20.828$\pm$0.226  &   20.894$\pm$0.283  &  2012-10-17  & 41.7642  &  41.1806 & 1.1117 & CAHA2.2    &(1)    \\
 120178      2003 OP$_{32 }$  &                   &   19.686$\pm$0.127  &  2011-10-31  & 41.6705  &  41.2996 & 1.2691 & CAHA2.2    &(1)    \\
             2003 QA$_{91 }$  & 23.790$\pm$1.163  &   23.158$\pm$0.933  &  2013-06-11  & 44.6206  &  44.3877 & 1.2746 &  INT       &(1,3)  \\
 120181      2003 UR$_{292}$  & 22.377$\pm$0.734  &   21.484$\pm$0.370  &  2012-09-19  & 26.7683  &  25.9857 & 1.3730 & CAHA2.2    &       \\
 143707      2003 UY$_{117}$  & 22.665$\pm$1.201  &   20.853$\pm$0.266  &  2011-10-31  & 32.8608  &  31.8758 & 0.2149 & CAHA2.2    &(1)    \\
 143707      2003 UY$_{117}$  & 20.503$\pm$0.201  &   20.614$\pm$0.258  &  2011-09-25  & 32.8502  &  32.0036 & 0.9558 & CAHA2.2    &       \\
 143707      2003 UY$_{117}$  &                   &   21.754$\pm$0.405  &  2012-09-17  & 32.9610  &  32.2100 & 1.1793 & CAHA2.2    &(1)    \\
  84922      2003 VS$_{2  }$  & 19.914$\pm$0.219  &   19.129$\pm$0.128  &  2012-12-09  & 36.5176  &  35.5619 & 0.3745 & CAHA2.2    &(1)    \\
  84922      2003 VS$_{2  }$  & 20.429$\pm$0.200  &   19.973$\pm$0.247  &  2012-10-17  & 36.5148  &  35.8418 & 1.1682 & CAHA2.2    &(1)    \\
 136204      2003 WL$_{7  }$  & 21.377$\pm$0.437  &   20.018$\pm$0.155  &  2011-10-31  & 14.9614  &  14.2362 & 2.6674 & CAHA2.2    &(1)    \\
 136204      2003 WL$_{7  }$  & 20.930$\pm$0.142  &   20.672$\pm$0.087  &  2012-10-15  & 15.0067  &  14.5833 & 3.5000 & CAHA2.2    &       \\
 120216      2004 EW$_{95 }$  & 21.131$\pm$0.181  &   20.513$\pm$0.144  &  2014-05-29  & 27.0955  &  26.2375 & 1.1547 & SOAR       &(1)    \\
  90568      2004 GV$_{9  }$  & 20.362$\pm$0.029  &   20.053$\pm$0.038  &  2014-04-02  & 39.3485  &  38.5014 & 0.7870 & SOAR       &(1)    \\
 307982      2004 PG$_{115}$  &                   &   20.183$\pm$0.247  &  2013-09-03  & 37.3300  &  36.3912 & 0.5687 &  OSN       &(1)    \\
 307982      2004 PG$_{115}$  & 21.321$\pm$0.212  &   20.414$\pm$0.097  &  2013-07-18  & 37.3077  &  36.4746 & 0.9090 & CAHA3.5    &       \\
             2004 PT$_{107}$  & 22.791$\pm$0.442  &   21.689$\pm$0.183  &  2012-10-16  & 38.2606  &  37.7248 & 1.2636 & CAHA2.2    &       \\
             2004 PT$_{107}$  & 21.906$\pm$0.284  &   20.994$\pm$0.179  &  2013-06-11  & 38.2517  &  37.8609 & 1.4142 &  INT       &(1)    \\
 144897      2004 UX$_{10 }$  & 19.474$\pm$0.106  &   19.632$\pm$0.116  &  2011-10-31  & 39.0037  &  38.0264 & 0.2566 & CAHA2.2    &(1)    \\
 144897      2004 UX$_{10 }$  & 20.534$\pm$0.083  &   19.905$\pm$0.052  &  2012-10-16  & 39.0466  &  38.0789 & 0.3597 & CAHA2.2    &       \\
 144897      2004 UX$_{10 }$  & 19.517$\pm$0.150  &   19.840$\pm$0.216  &  2011-09-25  & 38.9996  &  38.1471 & 0.7902 & CAHA2.2    &       \\   
 144897      2004 UX$_{10 }$  & 20.245$\pm$0.104  &   19.833$\pm$0.070  &  2011-09-24  & 38.9994  &  38.1563 & 0.8129 & CAHA2.2    &       \\   
 144897      2004 UX$_{10 }$  & 20.576$\pm$0.195  &   20.110$\pm$0.107  &  2012-09-19  & 39.0436  &  38.2544 & 0.9261 & CAHA2.2    &       \\
 230965      2004 XA$_{192}$  & 20.508$\pm$0.101  &   20.047$\pm$0.100  &  2012-12-11  & 35.6167  &  34.8167 & 0.9358 & CAHA2.2    &(3)    \\
 230965      2004 XA$_{192}$  & 20.110$\pm$0.144  &   19.718$\pm$0.089  &  2012-12-10  & 35.6169  &  34.8209 & 0.9453 & CAHA2.2    &(1,3)  \\
 230965      2004 XA$_{192}$  & 20.249$\pm$0.184  &   21.061$\pm$0.211  &  2012-12-08  & 35.6187  &  34.8313 & 0.9649 & CAHA2.2    &(3)    \\
 230965      2004 XA$_{192}$  & 20.468$\pm$0.218  &   19.638$\pm$0.109  &  2011-10-31  & 35.6725  &  35.2058 & 1.4196 & CAHA2.2    &(1,3)  \\
 230965      2004 XA$_{192}$  & 20.278$\pm$0.122  &   19.658$\pm$0.176  &  2012-09-16  & 35.6294  &  35.8013 & 1.5890 & CAHA2.2    &(3)    \\   
 303775      2005 QU$_{182}$  & 21.078$\pm$0.123  &   20.522$\pm$0.088  &  2013-07-18  & 50.1479  &  49.9291 & 1.1384 & CAHA3.5    &       \\
 145451      2005 RM$_{43 }$  & 20.033$\pm$0.165  &   19.691$\pm$0.127  &  2011-10-31  & 35.5824  &  34.6917 & 0.7200 & CAHA2.2    &(1)    \\
 145451      2005 RM$_{43 }$  & 19.981$\pm$0.123  &   19.808$\pm$0.063  &  2012-10-15  & 35.7194  &  34.9774 & 1.0843 & CAHA2.2    &       \\
 145451      2005 RM$_{43 }$  & 19.832$\pm$0.156  &   19.347$\pm$0.206  &  2011-09-25  & 35.5692  &  35.0655 & 1.4107 & CAHA2.2    &       \\   
 145451      2005 RM$_{43 }$  & 19.799$\pm$0.066  &   19.816$\pm$0.073  &  2011-09-24  & 35.5688  &  35.0795 & 1.4241 & CAHA2.2    &       \\   
 145452      2005 RN$_{43 }$  & 20.116$\pm$0.149  &   19.287$\pm$0.104  &  2013-09-02  & 40.6506  &  39.6582 & 0.2571 &  OSN       &(1)    \\
 145452      2005 RN$_{43 }$  & 19.940$\pm$0.117  &   19.398$\pm$0.172  &  2012-09-16  & 40.6611  &  39.7232 & 0.5098 & CAHA2.2    &       \\   
 145452      2005 RN$_{43 }$  & 20.100$\pm$0.235  &   19.414$\pm$0.109  &  2012-09-19  & 40.6610  &  39.7405 & 0.5668 & CAHA2.2    &       \\
 145452      2005 RN$_{43 }$  & 20.016$\pm$0.090  &   19.453$\pm$0.057  &  2011-09-24  & 40.6726  &  39.7937 & 0.6834 & CAHA2.2    &       \\   
 145452      2005 RN$_{43 }$  & 20.099$\pm$0.174  &   19.526$\pm$0.092  &  2011-10-31  & 40.6714  &  40.2388 & 1.2623 & CAHA2.2    &(1)    \\
 145453      2005 RR$_{43 }$  & 20.118$\pm$0.094  &   19.925$\pm$0.109  &  2012-12-11  & 39.0136  &  38.1198 & 0.6082 & CAHA2.2    &       \\
 145453      2005 RR$_{43 }$  & 20.365$\pm$0.220  &   19.547$\pm$0.101  &  2011-10-31  & 38.8984  &  37.9956 & 0.6200 & CAHA2.2    &(1)    \\
 145453      2005 RR$_{43 }$  & 20.067$\pm$0.075  &   19.828$\pm$0.071  &  2011-09-24  & 38.8883  &  38.3445 & 1.2540 & CAHA2.2    &       \\   
 145480      2005 TB$_{190}$  & 21.305$\pm$0.050  &   20.700$\pm$0.049  &  2013-09-03  & 46.2370  &  45.2423 & 0.2139 &  OSN       &(1)    \\
 145480      2005 TB$_{190}$  & 21.622$\pm$0.334  &   20.641$\pm$0.112  &  2013-07-17  & 46.2395  &  45.6431 & 1.0304 & CAHA3.5    &(1)    \\
 145480      2005 TB$_{190}$  & 22.596$\pm$0.648  &   21.191$\pm$0.290  &  2012-09-17  & 46.2640  &  45.2744 & 1.1793 & CAHA2.2    &(1)    \\
 145480      2005 TB$_{190}$  & 21.486$\pm$0.364  &   20.436$\pm$0.194  &  2012-12-08  & 46.2569  &  46.2518 & 1.2193 & CAHA2.2    &       \\
 145486      2005 UJ$_{438}$  & 20.981$\pm$0.470  &   20.918$\pm$0.460  &  2013-05-05  & 9.3817   &  9.0079  & 5.8384 &  OSN       &       \\
 145486      2005 UJ$_{438}$  & 21.517$\pm$1.266  &   20.113$\pm$0.367  &  2013-05-08  & 9.3875   &  9.0602  & 5.9318 &  OSN       &(1)    \\
 145486      2005 UJ$_{438}$  & 20.675$\pm$0.209  &   19.703$\pm$0.140  &  2012-12-08  & 9.1080   &  8.9316  & 6.1615 & CAHA2.2    &       \\
 202421      2005 UQ$_{513}$  & 21.388$\pm$0.342  &   20.421$\pm$0.217  &  2013-09-02  & 48.4570  &  47.7451 & 0.8554 &  OSN       &(1)    \\
 202421      2005 UQ$_{513}$  & 20.347$\pm$0.206  &   19.981$\pm$0.164  &  2012-12-09  & 48.5130  &  48.0673 & 1.0390 & CAHA2.2    &(1)    \\
             2007 OC$_{10 }$  & 20.994$\pm$0.185  &   20.435$\pm$0.094  &  2013-07-17  & 35.6259  &  34.7611 & 0.8738 & CAHA3.5    &(1)    \\
 225088      2007 OR$_{10 }$  & 21.358$\pm$0.476  &   20.904$\pm$0.665  &  2013-09-03  & 86.8923  &  85.8991 & 0.1105 &  OSN       &(1)    \\
 225088      2007 OR$_{10 }$  & 22.061$\pm$0.653  &   21.515$\pm$0.449  &  2012-09-17  & 86.6596  &  85.7407 & 0.2647 & CAHA2.2    &(1)    \\
 225088      2007 OR$_{10 }$  & 21.700$\pm$0.158  &                     &  2015-07-20  & 87.3397  &  86.5163 & 0.3975 & Live       &(1)    \\
 225088      2007 OR$_{10 }$  & 21.974$\pm$0.166  &                     &  2015-07-19  & 87.3391  &  86.5260 & 0.4068 & Live       &(1)    \\
 225088      2007 OR$_{10 }$  & 21.897$\pm$0.253  &   20.991$\pm$0.106  &  2013-07-17  & 86.8608  &  86.0473 & 0.4086 & CAHA3.5    &(1)    \\
 225088      2007 OR$_{10 }$  & 21.727$\pm$0.142  &                     &  2015-07-17  & 87.3381  &  86.5405 & 0.4199 & Live       &(1)    \\
 309239      2007 RW$_{10 }$  & 21.240$\pm$0.115  &   21.044$\pm$0.079  &  2013-07-17  & 28.3441  &  28.2058 & 2.0419 & CAHA3.5    &(1)    \\
 342842      2008 YB$_{3  }$  & 18.988$\pm$0.065  &   18.359$\pm$0.063  &  2013-04-16  & 7.3259   &  7.5578  & 7.5275 &  OSN       &       \\
             2013 AZ$_{60 }$  & 19.987$\pm$0.096  &   19.827$\pm$0.111  &  2013-04-15  & 8.6654   &  8.7576  & 6.5739 &  OSN       &(3)    \\
             2013 AZ$_{60 }$  & 20.188$\pm$0.136  &   20.019$\pm$0.127  &  2013-04-14  & 8.6676   &  8.7455  & 6.5840 &  OSN       &(3)    \\
  65489        Ceto           & 22.003$\pm$0.558  &   21.753$\pm$0.731  &  2013-05-11  & 34.1466  &  33.1959 & 0.5792 & CAHA3.5    &(1)    \\
  65489        Ceto           & 21.684$\pm$0.657  &   21.928$\pm$0.882  &  2013-07-21  & 34.3227  &  34.0991 & 1.6579 & CAHA3.5    &(1)    \\
  10199      Chariklo         & 18.812$\pm$0.183  &   18.321$\pm$0.158  &  2014-05-29  & 14.8053  &  13.8611 & 1.4804 & SOAR       &(1)    \\
  10199      Chariklo         &                   &   18.379$\pm$0.035  &  2014-05-22  & 14.8000  &  13.8959 & 1.8265 & OASI       &(1)    \\
  2060        Chiron          & 18.505$\pm$0.056  &   18.349$\pm$0.037  &  2012-10-16  & 17.3090  &  16.6359 & 2.4772 & CAHA2.2    &       \\
  2060        Chiron          & 18.263$\pm$0.132  &   17.843$\pm$0.080  &  2012-12-08  & 17.3610  &  17.5362 & 3.1833 & CAHA2.2    &       \\
 136108       Haumea          & 17.429$\pm$0.107  &   17.161$\pm$0.108  &  2013-05-06  & 50.8365  &  50.0445 & 0.7069 &  OSN       &(1,2)  \\
 136108       Haumea          & 17.580$\pm$0.078  &   17.250$\pm$0.073  &  2013-05-08  & 50.8362  &  50.0582 & 0.7273 &  OSN       &(1)    \\
 136472      Makemake         & 16.966$\pm$0.104  &   16.421$\pm$0.105  &  2013-05-06  & 52.3120  &  51.7204 & 0.8969 &  OSN       &(1,2)  \\
 136472      Makemake         &                   &   16.469$\pm$0.091  &  2013-05-07  & 52.3122  &  51.7330 & 0.9070 &  OSN       &(1)    \\
 136472      Makemake         & 17.241$\pm$0.070  &   16.796$\pm$0.060  &  2013-05-08  & 52.3123  &  51.7449 & 0.9162 &  OSN       &(1)    \\
  5145        Pholus          & 21.677$\pm$0.075  &   20.911$\pm$0.035  &  2013-06-10  & 25.2536  &  24.2802 & 0.6679 &  INT       &       \\
  5145        Pholus          & 21.765$\pm$0.181  &   21.221$\pm$0.190  &  2013-06-11  & 25.2552  &  24.2817 & 0.6685 &  INT       &(1)    \\
 120347      Salacia          & 20.482$\pm$0.121  &   20.212$\pm$0.086  &  2011-09-24  & 44.2535  &  43.3414 & 0.5446 & CAHA2.2    &       \\   
 120347      Salacia          & 20.558$\pm$0.231  &   20.713$\pm$0.326  &  2011-09-25  & 44.2537  &  43.3440 & 0.5505 & CAHA2.2    &       \\
 120347      Salacia          & 20.800$\pm$0.146  &   20.309$\pm$0.076  &  2012-10-15  & 44.3415  &  43.5369 & 0.7637 & CAHA2.2    &       \\
 120347      Salacia          & 20.795$\pm$0.267  &   20.128$\pm$0.146  &  2011-10-31  & 44.2619  &  43.6166 & 0.9799 & CAHA2.2    &(1)    \\
  88611   Teharonhiawako      & 22.538$\pm$0.517  &   22.609$\pm$0.623  &  2012-09-16  & 45.1115  &  44.1527 & 0.3812 & CAHA2.2    &(3)    \\
  42355       Typhon          & 20.504$\pm$0.358  &   19.920$\pm$0.365  &  2013-05-11  & 19.0989  &  18.3380 & 2.0288 & CAHA3.5    &(1)    \\
 174567       Varda           & 20.387$\pm$0.057  &   19.722$\pm$0.060  &  2013-06-03  & 47.3363  &  46.3863 & 0.4357 & CAHA3.5    &       \\
 174567       Varda           & 20.517$\pm$0.290  &   19.948$\pm$0.616  &  2013-05-10  & 47.3444  &  46.4795 & 0.6414 & CAHA3.5    &       \\
 174567       Varda           & 20.474$\pm$0.150  &   20.235$\pm$0.132  &  2013-05-05  & 47.3456  &  46.5173 & 0.7075 &  OSN       &       \\
 174567       Varda           & 20.116$\pm$0.084  &   19.918$\pm$0.089  &  2013-04-14  & 47.3528  &  46.7400 & 0.9715 &  OSN       &       \\
\end{longtable}
\tablefoot{
(1) Average ext. coeff.
(2) Average zero points.
(3) Never reported before.
}
\end{longtab}

\addtocounter{table}{-1}
\begin{longtab}
\begin{landscape}
\begin{longtable}{r c c c c l}
\caption{\label{table:1}Absolute Magnitudes}\\
\hline\hline
Object & \hv & $\beta$ (mag per degree) & N &  $\Delta m$&References \\    
\hline
\endfirsthead
\caption{continued.}\\
\hline\hline
Object & \hv & $\beta$ (mag per degree) & N &  $\Delta m$&References \\    
\hline
\endhead
\hline
\endfoot
   15760     1992 QB$_{1  }$    &$ 7.839\pm0.097$&$  -0.193\pm0.132$&   3& -- &{ TR00,JL01,Bo01}\\
   15788     1993 SB            &$ 7.995\pm0.059$&$   0.374\pm0.066$&   5& -- &{ Da00,TR00,GH01,JL01,De01}\\
   15789     1993 SC            &$ 7.393\pm0.020$&$   0.050\pm0.017$&   8&0.04&{ JL98,Da00,Te97,JL01,TR97}\\
             1994 EV$_{3  }$    &$ 8.183\pm0.247$&$  -0.803\pm0.329$&   3& -- &{ Bo02,Bo01,GH01}\\
   16684     1994 JQ$_{1  }$    &$ 7.031\pm0.078$&$   0.570\pm0.125$&   5& -- &{ Bo02,TR03,GH01}\\
   15820     1994 TB            &$ 8.017\pm0.226$&$   0.133\pm0.152$&   9&0.34&{ Da00,RT99,De01,JL01,TR97}\\
   19255     1994 VK$_{8  }$    &$ 7.840\pm0.923$&$  -0.173\pm0.976$&   3&0.42&{ TR00,Do01,RT99}\\
             1995 HM$_{5  }$    &$ 8.315\pm0.100$&$   0.037\pm0.074$&   5& -- &{ RT99,GH01,Ba00}\\
   32929     1995 QY$_{9  }$    &$ 8.136\pm0.515$&$  -0.108\pm0.459$&   4&0.60&{ Da00,RT99,GH01}\\
   24835     1995 SM$_{55 }$    &$ 4.584\pm0.178$&$   0.139\pm0.198$&   8&0.19&{ TR03,MB03,GH01,De01,Bo01,Do02,{\bf TW}}\\
   26181     1996 GQ$_{21 }$    &$ 5.073\pm0.050$&$   0.858\pm0.124$&   6&0.10&{ MB03,Bo02,{\bf TW}}\\
             1996 RQ$_{20 }$    &$ 7.201\pm0.073$&$  -0.065\pm0.075$&   5& -- &{ RT99,De01,JL01,Sn10,Bo01}\\
             1996 RR$_{20 }$    &$ 6.986\pm0.128$&$   0.391\pm0.210$&   3& -- &{ Bo02,TR00,JL01}\\
   19299     1996 SZ$_{4  }$    &$ 8.564\pm0.034$&$   0.307\pm0.054$&   4& -- &{ TR00,Da00,JL01,Bo02}\\
             1996 TK$_{66 }$    &$ 7.031\pm0.086$&$  -0.280\pm0.115$&   3& -- &{ TR00,JL01,Do02}\\
   15874     1996 TL$_{66 }$    &$ 5.257\pm0.100$&$   0.375\pm0.112$&   5&0.12&{ JL01,RT99,JL98,Da00,Bo01}\\
   19308     1996 TO$_{66 }$    &$ 4.806\pm0.144$&$   0.150\pm0.197$&   7&0.33&{ JL98,Da00,RT99,GH01,Sh10,JL01,Bo01}\\
   15875     1996 TP$_{66 }$    &$ 7.461\pm0.084$&$   0.127\pm0.072$&   5&0.12&{ RT99,JL01,JL98,Da00,Bo01}\\
  118228     1996 TQ$_{66 }$    &$ 8.006\pm0.422$&$  -0.415\pm0.680$&   4&0.22&{ RT99,JL01,GH01,Da00}\\
             1996 TS$_{66 }$    &$ 6.535\pm0.167$&$   0.083\pm0.220$&   4&0.16&{ JL01,RT99,JL98,Da00}\\
   33001     1997 CU$_{29 }$    &$ 6.808\pm0.057$&$   0.075\pm0.087$&   4& -- &{ Do01,TR00,Ba00,JL01}\\
             1997 QH$_{4  }$    &$ 7.216\pm0.143$&$   0.451\pm0.142$&   4& -- &{ TR00,JL01,Bo02,De01}\\
   24952     1997 QJ$_{4  }$    &$ 7.754\pm0.113$&$   0.290\pm0.103$&   5& -- &{ De01,GH01,Da00,JL01,Bo02}\\
   91133     1998 HK$_{151}$    &$ 7.340\pm0.056$&$   0.127\pm0.088$&   5&0.15&{ Bo01,Do01,MB03,Do02}\\
  385194     1998 KG$_{62 }$    &$ 7.647\pm0.194$&$  -0.748\pm0.205$&   3& -- &{ Bo02,GH01,Do01}\\
   26308     1998 SM$_{165}$    &$ 5.938\pm0.363$&$   0.446\pm0.376$&   3&0.56&{ MB03,TR00,De01}\\
   35671     1998 SN$_{165}$    &$ 5.879\pm0.109$&$  -0.031\pm0.115$&   6&0.16&{ De01,MB03,Do01,Fo04,GH01,JL01}\\
             1998 UR$_{43 }$    &$ 9.047\pm0.108$&$  -0.764\pm0.165$&   3& -- &{ GH01,De01}\\
   33340     1998 VG$_{44 }$    &$ 6.599\pm0.205$&$   0.228\pm0.158$&   3&0.10&{ Do02,Bo01,Do01}\\
             1999 CD$_{158}$    &$ 5.289\pm0.092$&$   0.092\pm0.119$&   3& -- &{ Do02,Sn10,De01}\\
   26375     1999 DE$_{9  }$    &$ 5.120\pm0.024$&$   0.183\pm0.032$&  36&0.10&{ Ra07,DM09,MB03,Te03,Do02,JL01,De01}\\
             1999 HS$_{11 }$    &$ 6.843\pm0.555$&$   0.233\pm0.728$&   3& -- &{ Px04,TR03,Do01}\\
   40314     1999 KR$_{16 }$    &$ 6.316\pm0.139$&$  -0.126\pm0.180$&   4&0.18&{ Bo02,JL01,{\bf TW}}\\
   44594     1999 OX$_{3  }$    &$ 7.980\pm0.092$&$  -0.086\pm0.057$&  12&0.11&{ Do01,Do02,De01,Pe10,TR00,Bo14,Do05,MB03,Px04,Sh10}\\
   86047     1999 OY$_{3  }$    &$ 6.579\pm0.044$&$   0.067\pm0.042$&   4& -- &{ Do02,TR00,Sn10,Bo02}\\
   86177     1999 RY$_{215}$    &$ 7.097\pm0.084$&$   0.341\pm0.111$&   3& -- &{ Sn10,Bo02,Do01}\\
   47171     1999 TC$_{36 }$    &$ 5.395\pm0.030$&$   0.111\pm0.027$&  45&0.07&{ Ra07,Te03,MB03,De01,Do03,DM09,Do01,Bo01,{\bf TW}}\\
   29981     1999 TD$_{10 }$    &$ 9.105\pm0.430$&$   0.033\pm0.122$&  21&0.65&{ MB03,Ra07,De01,TR03,Do02}\\
   47932     2000 GN$_{171}$    &$ 6.776\pm0.243$&$  -0.101\pm0.186$&  29&0.61&{ Ra07,MB03,Bo02,DM09,{\bf TW}}\\
  138537     2000 OK$_{67 }$    &$ 6.629\pm0.694$&$   0.089\pm0.518$&   3& -- &{ Do02,De01}\\
   82075     2000 YW$_{134}$    &$ 4.378\pm0.687$&$   0.377\pm0.552$&   3&0.10&{ SS09,Do05,{\bf TW}}\\
   82158     2001 FP$_{185}$    &$ 6.420\pm0.062$&$   0.123\pm0.052$&   5&0.06&{ Te03,Px04,Do05,{\bf TW}}\\
             2001 KA$_{77 }$    &$ 5.646\pm0.090$&$   0.130\pm0.095$&   3& -- &{ Do05,Px04,Do02}\\
             2001 KD$_{77 }$    &$ 6.299\pm0.099$&$   0.141\pm0.082$&   3&0.07&{ Px04,Do02,{\bf TW}}\\
   42301     2001 UR$_{163}$    &$ 4.529\pm0.063$&$   0.364\pm0.117$&   3&0.08&{ Do05,SS09,Pe10}\\
   55565     2002 AW$_{197}$    &$ 3.593\pm0.023$&$   0.206\pm0.029$&  39&0.04&{ Ra07,DM09,Fo04,{\bf TW}}\\
             2002 GP$_{32 }$    &$ 7.133\pm0.027$&$  -0.135\pm0.036$&   4&0.03&{ Do05,{\bf TW}}\\
   95626     2002 GZ$_{32 }$    &$ 7.419\pm0.126$&$   0.043\pm0.064$&  29&0.15&{ Ra07,Fo04,Do05,Te03,{\bf TW}}\\
  119951     2002 KX$_{14 }$    &$ 4.978\pm0.017$&$   0.114\pm0.031$&  20& -- &{ Ra07,Ro10,Bo14,DM09,{\bf TW}}\\
  250112     2002 KY$_{14 }$    &$11.808\pm0.763$&$  -0.274\pm0.193$&   4&0.13&{ Bo14,Pe10,{\bf TW}}\\
   73480     2002 PN$_{34 }$    &$ 8.618\pm0.054$&$   0.090\pm0.027$&  57&0.18&{ Ra07,Pe10,Te03}\\
   55636     2002 TX$_{300}$    &$ 3.574\pm0.055$&$   0.005\pm0.044$&  37&0.09&{ Ra07,Te03,Or04,Do05}\\
   55637     2002 UX$_{25 }$    &$ 3.883\pm0.048$&$   0.159\pm0.056$&  42&0.21&{ Ra07,SS09,DM09,{\bf TW}}\\
   55638     2002 VE$_{95 }$    &$ 5.813\pm0.037$&$   0.089\pm0.024$&  43&0.08&{ Pe10,Ra07,{\bf TW}}\\
  127546     2002 XU$_{93 }$    &$ 7.031\pm0.859$&$   0.498\pm0.320$&   5& -- &{ Sh10,{\bf TW}}\\
  208996     2003 AZ$_{84 }$    &$ 3.779\pm0.114$&$   0.074\pm0.118$&   5&0.14&{ DM09,Pe10,Fo04,SS09,Bo14}\\
  120061     2003 CO$_{1  }$    &$ 9.146\pm0.056$&$   0.092\pm0.015$&   5&0.07&{ Pe10,Pe13,Te03}\\
  133067     2003 FB$_{128}$    &$ 6.922\pm0.566$&$   0.422\pm0.469$&   3& -- &{ Pe13,Bo14}\\
             2003 FE$_{128}$    &$ 7.381\pm0.256$&$  -0.349\pm0.209$&   5& -- &{ Pe13,Bo14}\\
  120132     2003 FY$_{128}$    &$ 4.632\pm0.187$&$   0.535\pm0.145$&   7&0.15&{ Pe10,Sh10,DM09,Bo14,{\bf TW}}\\
  385437     2003 GH$_{55 }$    &$ 7.319\pm0.247$&$  -0.880\pm0.251$&   3& -- &{ Pe13,Bo14}\\
  120178     2003 OP$_{32 }$    &$ 4.067\pm0.318$&$   0.045\pm0.280$&  11&0.26&{ Pe10,Bo14,Pe13,{\bf TW}}\\
  143707     2003 UY$_{117}$    &$ 5.830\pm1.299$&$  -0.230\pm1.329$&   3& -- &{ {\bf TW}}\\
             2003 UZ$_{117}$    &$ 5.185\pm0.054$&$   0.214\pm0.073$&   3& -- &{ Pe10,Bo14,DM09}\\
             2003 UZ$_{413}$    &$ 4.361\pm0.068$&$   0.144\pm0.096$&   3& -- &{ Pe10}\\
  136204     2003 WL$_{7  }$    &$ 8.897\pm0.149$&$   0.089\pm0.049$&   4&0.05&{ Pe13,{\bf TW}}\\
  120216     2004 EW$_{95 }$    &$ 6.579\pm0.021$&$   0.071\pm0.024$&   4& -- &{ Bo14,Pe13,{\bf TW}}\\
   90568     2004 GV$_{9  }$    &$ 3.409\pm0.357$&$   1.353\pm0.542$&   3&0.16&{ Bo14,DM09,{\bf TW}}\\
  307982     2004 PG$_{115}$    &$ 4.874\pm0.064$&$   0.505\pm0.051$&   8& -- &{ Pe13,Bo14,{\bf TW}}\\
  120348     2004 TY$_{364}$    &$ 4.519\pm0.137$&$   0.146\pm0.103$&  32&0.22&{ Ra07,Pe10}\\
  144897     2004 UX$_{10 }$    &$ 4.825\pm0.097$&$   0.061\pm0.103$&   8&0.08&{ Ro10,Pe10,{\bf TW}}\\
  230965     2004 XA$_{192}$    &$ 5.059\pm0.085$&$  -0.175\pm0.070$&   5&0.07&{ {\bf TW}}\\
  303775     2005 QU$_{182}$    &$ 3.853\pm0.028$&$   0.277\pm0.034$&   5& -- &{ Pe13,Bo14,{\bf TW}}\\
  145451     2005 RM$_{43 }$    &$ 4.704\pm0.081$&$  -0.028\pm0.064$&   6&0.04&{ Bo14,DM09,{\bf TW}}\\
  145452     2005 RN$_{43 }$    &$ 3.882\pm0.036$&$   0.138\pm0.030$&  10&0.04&{ DM09,Pe13,{\bf TW}}\\
  145453     2005 RR$_{43 }$    &$ 4.252\pm0.067$&$  -0.003\pm0.065$&   5&0.06&{ Pe10,DM09,{\bf TW}}\\
  145480     2005 TB$_{190}$    &$ 4.676\pm0.084$&$   0.052\pm0.106$&   7&0.12&{ Bo14,Pe13,{\bf TW}}\\
  145486     2005 UJ$_{438}$    &$14.602\pm0.617$&$  -0.412\pm0.098$&   5&0.13&{ Bo14,Pe13,{\bf TW}}\\
             2007 OC$_{10 }$    &$ 5.330\pm0.825$&$   0.223\pm0.740$&   3& -- &{ Pe13,{\bf TW}}\\
  225088     2007 OR$_{10 }$    &$ 2.316\pm0.124$&$   0.257\pm0.505$&   7& -- &{ Bo14,{\bf TW}}\\
  281371     2008 FC$_{76 }$    &$ 9.486\pm0.078$&$   0.101\pm0.016$&   4& -- &{ Pe10,Px12,Pe13}\\
  342842     2008 YB$_{3  }$    &$11.024\pm0.696$&$  -0.104\pm0.095$&   3&0.20&{ Pa13,Sh10,{\bf TW}}\\
   55576      Amycus            &$ 8.213\pm0.621$&$   0.052\pm0.351$&   3&0.16&{ Px04,Fo04,Pe10}\\
   8405      Asbolus            &$ 9.138\pm0.130$&$   0.042\pm0.029$&  43&0.55&{ Ro97,BL97,Ra07,RM02}\\
   54598      Bienor            &$ 7.656\pm0.443$&$   0.130\pm0.170$&  57&0.75&{ Ra07,DM09,Do02,Te03,De01}\\
   66652     Borasisi           &$ 6.032\pm0.040$&$   0.231\pm0.062$&   3&0.05&{ Do01,MB03}\\
   65489       Ceto             &$ 6.573\pm0.126$&$   0.196\pm0.096$&   8&0.13&{ Bo14,Te03,Pe13,{\bf TW}}\\
   19521      Chaos             &$ 4.987\pm0.065$&$   0.102\pm0.070$&   6&0.10&{ De01,TR00,Do02,Bo01,Da00,Ba00}\\
   10199     Chariklo           &$ 6.870\pm0.055$&$   0.064\pm0.016$&  22&0.10&{ Pe01,Be10,DM09,Fo14,RT99,MB99,JL01,Pe10,{\bf TW}}\\
   2060       Chiron            &$ 6.399\pm0.019$&$   0.083\pm0.005$&  37&0.09&{ Be10,Du02,Je02,{\bf TW}}\\
   83982     Crantor            &$ 9.096\pm0.405$&$   0.110\pm0.149$&   5&0.34&{ Px04,DM09,Fo04,Te03}\\
   52975     Cyllarus           &$ 9.064\pm0.041$&$   0.184\pm0.029$&   4& -- &{ De01,Te03,Do02,Bo01}\\
   31824      Elatus            &$10.592\pm0.171$&$   0.078\pm0.031$&   6&0.24&{ TR03,De01,Do02,Pe01}\\
  136199       Eris             &$-1.124\pm0.025$&$   0.119\pm0.056$&  79&0.10&{ DM09,Ra07,Ca06,}\\
   38628       Huya             &$ 4.975\pm0.037$&$   0.173\pm0.026$&  98&0.10&{ Fe01,SR02,TR03,Bo02,MB03,Do01,JL01}\\
   28978      Ixion             &$ 3.774\pm0.021$&$   0.194\pm0.031$&  40&0.05&{ Ra07,DM09,Do02}\\
   58534      Logos             &$ 7.411\pm0.041$&$   0.055\pm0.057$&   5& -- &{ Ba00,GH01,JL01,Bo01}\\
  136472     Makemake           &$ 0.009\pm0.012$&$   0.202\pm0.015$&  55&0.03&{ Ra07,{\bf TW}}\\
   52872     Okyrhoe            &$11.441\pm0.062$&$  -0.023\pm0.017$&   6&0.07&{ Do03,De01,Pe10,TR03,Bo14,Do01}\\
   90482      Orcus             &$ 2.280\pm0.021$&$   0.160\pm0.022$&  30&0.04&{ Ra07,Pe10,dB05}\\
   49036      Pelion            &$10.911\pm0.069$&$  -0.074\pm0.026$&   3& -- &{ Do02,Bo02,TR00}\\
   5145       Pholus            &$ 7.474\pm0.309$&$   0.153\pm0.156$&  15&0.60&{ Mu92,BB92,Be10,Ro97,{\bf TW}}\\
   50000      Quaoar            &$ 2.777\pm0.250$&$   0.117\pm0.221$&  45&0.30&{ Ra07,Te03,DM09,Fo04}\\
  120347     Salacia            &$ 4.151\pm0.030$&$   0.132\pm0.028$&   9&0.03&{ Sn10,Bo14,Pe13,{\bf TW}}\\
   90377      Sedna             &$ 1.669\pm0.004$&$   0.266\pm0.008$& 170&0.02&{ Ra07,Ba05,Sh10,Pe10}\\
   79360    Sila-Nunam          &$ 5.573\pm0.224$&$   0.095\pm0.209$&   6&0.22&{ Da00,Bo01,Ba00,JL01,RT99}\\
   32532     Thereus            &$ 9.454\pm0.137$&$   0.061\pm0.034$&  67&0.34&{ Ra07,Te03,FD03,DM09,Ba02}\\
   42355      Typhon            &$ 7.670\pm0.026$&$   0.128\pm0.013$&  22&0.07&{ Ra07,Te03,DM09,Pe10,Px04,{\bf TW}}\\
  174567      Varda             &$ 3.988\pm0.048$&$  -0.455\pm0.071$&   9&0.06&{ Bo14,Pe13,Pe10,{\bf TW}}\\
   20000      Varuna            &$ 3.966\pm0.233$&$   0.104\pm0.246$&  30&0.50&{ Ra07,Do02,Pe10,JS02,TR03}\\
\end{longtable}
\tablefoot{
BB92 = Buie and Bus (1992),
Mu92 = Mueller et al. (1992),
BL97 = Brown and Luu (1997),
TR97 = Tegler and Romanishin (1997),
Te97 = Tegler et al. (1997),
Ro97 = Romanishin et al. (1997),
JL98 = Jewitt and Luu (1998),
MB99 = McBride et al. (1999),
RT99 = Romanishin and Tegler (1999),
Ba00 = Barucci et al. (2000),
Da00 = Davies et al. (2000),
TR00 = Tegler and Romanishin (2000),	
Bo01 = Boehnhardt et al. (2001), 
De01 = Delsanti et al. (2001), 
Do01 = Doressoundiram et al. (2001),
Fe01 = Ferrin et al. (2001),
GH01 = Gil-Hutton and Licandro (2001),
JL01 = Jewitt and Luu (2001),	
Pe01 = Peixinho et al. (2001),
Ba02 = Barucci et al. (2002),
Bo02 = Boehnhardt et al. (2002),
Do02 = Doressoundiram et al. (2002), 
Du02 = Duffard et al. (2002),
Je02 = Jewitt (2002).
JS02 = Jewitt and Sheppard (2002),
RM02 = Romon-Martin et al. (2002),
SR02 = Schaefer and Rabinowitz (2002),
FD03 = Farnham and Davies (2003),
MB03 = McBride et al. (2003),
Do03 = Dotto et al. (2003),
TR03 = Tegler and Romanishin (2003),
Te03 = Tegler et al. (2003),
Fo04 = Fornasier et al. (2004),
Or04 = Ortiz et al. (2004),
Px04 = Peixinho et al. (2004),
Ba05 = Barucci et al. (2005),
dB05 = de Bergh et al. (2005),
Do05 = Doressoundiram et al. (2005),
Ca06 = Carraro et al. (2006),
Ra07 = Rabinowitz et al. (2007),
DM09 = DeMeo et al. (2009),
SS09 = Santos-Sanz et al. (2009),
Be10 = Belskaya et al. (2010),
Pe10 = Perna et al. (2010),
Ro10 = Romanishin et al. (2010), 
Sh10 = Sheppard (2010),
Sn10 = Snodgrass et al. (2010),
Px12 = Peixinho et al. (2012),
Pe13 = Perna et al. (2013),
PA13 = Pinilla-Alonso et al. (2013),
Bo14 = B\"ohnhardt et al. (2014),
Fo14 = Fornasier et al. (2014),
{\bf TW} = This work.
}
\end{landscape}
\end{longtab}

\addtocounter{table}{-1}
\begin{longtab}
\begin{landscape}
\begin{longtable}{r|cc|cc|cc|c}
\caption{\label{table:comp}Comparison between the values of \hv~and $\beta$ from this work with those from three selected references}\\
\hline\hline
& This work && Ra07 &&Pe13&&Bo13\\
\hline
Object       & \hv & $\beta$ (mag per degree) & \hv & $\beta$ (mag per degree)  & \hv & $\beta$ (mag per degree) & \hv \\
\hline
\endfirsthead
\caption{continued.}\\
\hline\hline
& This work && Ra07 &&Pe13&&Bo13\\
\hline
Object       & \hv & $\beta$ (mag per degree) & \hv & $\beta$ (mag per degree)  & \hv & $\beta$ (mag per degree) & \hv \\
\hline
\endhead
\hline
\endfoot
    1999 DE$_{9  }$   &$ 5.120\pm 0.024$&$  0.183\pm 0.032$&     5.103 $\pm$ 0.029 &     0.209 $\pm$ 0.035 &                     &               &                     \\
    1999 TC$_{36 }$   &$ 5.395\pm 0.030$&$  0.111\pm 0.027$&     5.272 $\pm$ 0.055 &     0.131 $\pm$ 0.049 &                     &               &                     \\
    1999 TD$_{10 }$   &$ 9.105\pm 0.430$&$  0.033\pm 0.122$&     8.793 $\pm$ 0.029 &     0.150 $\pm$ 0.014 &                     &               &                     \\
    2000 GN$_{171}$   &$ 6.776\pm 0.243$&$ -0.101\pm 0.186$&     6.368 $\pm$ 0.034 &     0.143 $\pm$ 0.030 &                     &               &                     \\
    2002 AW$_{197}$   &$ 3.593\pm 0.023$&$  0.206\pm 0.029$&     3.568 $\pm$ 0.030 &     0.128 $\pm$ 0.040 &                     &               &                     \\
    2002 GZ$_{32 }$   &$ 7.419\pm 0.126$&$  0.043\pm 0.064$&     7.389 $\pm$ 0.059 &    -0.025 $\pm$ 0.041 &                     &               &                     \\
    2002 KX$_{14 }$   &$ 4.978\pm 0.017$&$  0.114\pm 0.031$&     4.862 $\pm$ 0.038 &     0.159 $\pm$ 0.044 &                     &               &     5.07 $\pm$ 0.03 \\
    2002 KY$_{14 }$   &$11.808\pm 0.763$&$ -0.274\pm 0.193$&                       &                       &                     &               &    10.50 $\pm$ 0.08 \\
    2002 PN$_{34 }$   &$ 8.618\pm 0.054$&$  0.090\pm 0.027$&     8.660 $\pm$ 0.017 &     0.043 $\pm$ 0.005 &                     &               &                     \\
    2002 TX$_{300}$   &$ 3.574\pm 0.055$&$  0.005\pm 0.044$&     3.365 $\pm$ 0.044 &     0.158 $\pm$ 0.053 &                     &               &                     \\
    2002 UX$_{25 }$   &$ 3.883\pm 0.048$&$  0.159\pm 0.056$&     3.873 $\pm$ 0.020 &     0.158 $\pm$ 0.025 &                     &               &                     \\
    2002 VE$_{95 }$   &$ 5.813\pm 0.037$&$  0.089\pm 0.024$&     5.748 $\pm$ 0.058 &     0.121 $\pm$ 0.039 &                     &               &                     \\
    2003 AZ$_{84 }$   &$ 3.779\pm 0.114$&$  0.074\pm 0.118$&                       &                       &                     &               &     3.54 $\pm$ 0.03 \\
    2003 CO$_{1  }$   &$ 9.146\pm 0.056$&$  0.092\pm 0.015$&                       &                       &     9.07 $\pm$ 0.05 &$ 0.09\pm 0.01$&                     \\
    2003 FB$_{128}$   &$ 6.922\pm 0.566$&$  0.422\pm 0.469$&                       &                       &     7.09 $\pm$ 0.20 &               &     7.26 $\pm$ 0.05 \\
    2003 FE$_{128}$   &$ 7.381\pm 0.256$&$ -0.349\pm 0.209$&                       &                       &     6.74 $\pm$ 0.18 &               &     6.94 $\pm$ 0.07 \\
    2003 FY$_{128}$   &$ 4.632\pm 0.187$&$  0.535\pm 0.145$&                       &                       &                     &               &     5.36 $\pm$ 0.08 \\
    2003 GH$_{55 }$   &$ 7.319\pm 0.247$&$ -0.880\pm 0.251$&                       &                       &     6.32 $\pm$ 0.13 &               &{\bf 6.18 $\pm$ 0.04}\\
    2003 OP$_{32 }$   &$ 4.067\pm 0.318$&$  0.045\pm 0.280$&                       &                       &     3.99 $\pm$ 0.11 &$ 0.05\pm 0.08$&     3.79 $\pm$ 0.08 \\
    2003 UZ$_{117}$   &$ 5.185\pm 0.054$&$  0.214\pm 0.073$&                       &                       &                     &               &     5.27 $\pm$ 0.02 \\
    2003 WL$_{7  }$   &$ 8.897\pm 0.149$&$  0.089\pm 0.049$&                       &                       &     8.75 $\pm$ 0.16 &               &                     \\
    2004 EW$_{95 }$   &$ 6.579\pm 0.021$&$  0.071\pm 0.024$&                       &                       &     6.39 $\pm$ 0.15 &               &     6.52 $\pm$ 0.01 \\
    2004 GV$_{9  }$   &$ 3.409\pm 0.357$&$  1.353\pm 0.542$&                       &                       &                     &               &     4.03 $\pm$ 0.03 \\
    2004 PG$_{115}$   &$ 4.874\pm 0.064$&$  0.505\pm 0.051$&                       &                       &     5.23 $\pm$ 0.15 &               &{\bf 5.53 $\pm$ 0.05}\\
    2004 TY$_{364}$   &$ 4.519\pm 0.137$&$  0.146\pm 0.103$&     4.434 $\pm$ 0.074 &     0.184 $\pm$ 0.070 &                     &               &                     \\
    2005 QU$_{182}$   &$ 3.853\pm 0.028$&$  0.277\pm 0.034$&                       &                       &     3.82 $\pm$ 0.12 &               &     3.99 $\pm$ 0.02 \\
    2005 RM$_{43 }$   &$ 4.704\pm 0.081$&$ -0.028\pm 0.064$&                       &                       &                     &               &     4.52 $\pm$ 0.01 \\
    2005 RN$_{43 }$   &$ 3.882\pm 0.036$&$  0.138\pm 0.030$&                       &                       &     3.72 $\pm$ 0.05 &$ 0.18\pm 0.03$&                     \\
    2005 TB$_{190}$   &$ 4.676\pm 0.084$&$  0.052\pm 0.106$&                       &                       &     4.62 $\pm$ 0.15 &               &     4.56 $\pm$ 0.02 \\
    2005 UJ$_{438}$   &$14.602\pm 0.617$&$ -0.412\pm 0.098$&                       &                       &{\bf11.14 $\pm$ 0.32}&               &                     \\
    2007 OC$_{10 }$   &$ 5.330\pm 0.825$&$  0.223\pm 0.740$&                       &                       &     5.36 $\pm$ 0.13 &               &                     \\
    2007 OR$_{10 }$   &$ 2.316\pm 0.124$&$  0.257\pm 0.505$&                       &                       &                     &               &     2.34 $\pm$ 0.01 \\
    2008 FC$_{76 }$   &$ 9.486\pm 0.078$&$  0.101\pm 0.016$&                       &                       &     9.43 $\pm$ 0.13 &$ 0.09\pm 0.02$&                     \\
     Asbolus          &$ 9.138\pm 0.130$&$  0.042\pm 0.029$&     9.107 $\pm$ 0.016 &     0.050 $\pm$ 0.004 &                     &               &                     \\
      Bienor          &$ 7.656\pm 0.443$&$  0.130\pm 0.170$&     7.588 $\pm$ 0.035 &     0.095 $\pm$ 0.016 &                     &               &                     \\
       Ceto           &$ 6.573\pm 0.126$&$  0.196\pm 0.096$&                       &                       &     6.58 $\pm$ 0.10 &$ 0.10\pm 0.06$&     6.60 $\pm$ 0.01 \\
       Eris           &$-1.124\pm 0.025$&$  0.119\pm 0.056$&    -1.116 $\pm$ 0.009 &     0.105 $\pm$ 0.020 &                     &               &                     \\
       Huya           &$ 5.015\pm 0.021$&$  0.124\pm 0.015$&     5.048 $\pm$ 0.005 &     0.155 $\pm$ 0.041 &                     &               &                     \\
      Ixion           &$ 3.774\pm 0.021$&$  0.194\pm 0.031$&     3.766 $\pm$ 0.042 &     0.133 $\pm$ 0.043 &                     &               &                     \\
     Makemake         &$ 0.009\pm 0.012$&$  0.202\pm 0.015$&{\bf 0.091 $\pm$ 0.015}&{\bf 0.054 $\pm$ 0.019}&                     &               &                     \\
     Okyrhoe          &$11.441\pm 0.062$&$ -0.023\pm 0.017$&                       &                       &                     &               &{\bf10.83 $\pm$ 0.01}\\
      Orcus           &$ 2.280\pm 0.021$&$  0.160\pm 0.022$&     2.328 $\pm$ 0.028 &     0.114 $\pm$ 0.030 &                     &               &                     \\
      Quaoar          &$ 2.777\pm 0.250$&$  0.117\pm 0.221$&     2.729 $\pm$ 0.025 &     0.159 $\pm$ 0.027 &                     &               &                     \\
     Salacia          &$ 4.151\pm 0.030$&$  0.132\pm 0.028$&                       &                       &     4.26 $\pm$ 0.06 &$ 0.04\pm 0.06$&     4.01 $\pm$ 0.02 \\
      Sedna           &$ 1.443\pm 0.003$&$  0.200\pm 0.006$&{\bf 1.829 $\pm$ 0.048}&                       &                     &               &                     \\
     Thereus          &$ 9.454\pm 0.137$&$  0.061\pm 0.034$&     9.417 $\pm$ 0.014 &     0.072 $\pm$ 0.004 &                     &               &                     \\
      Typhon          &$ 7.670\pm 0.026$&$  0.128\pm 0.013$&     7.676 $\pm$ 0.037 &     0.126 $\pm$ 0.022 &                     &               &                     \\
      Varda           &$ 3.988\pm 0.048$&$ -0.455\pm 0.071$&                       &                       &{\bf 3.51 $\pm$ 0.06}&               &     3.93 $\pm$ 0.07 \\
      Varuna          &$ 3.966\pm 0.233$&$  0.104\pm 0.246$&     3.760 $\pm$ 0.032 &     0.278 $\pm$ 0.047 &                     &               &                     \\
\end{longtable}
\end{landscape}
\end{longtab}


\end{appendix}
\end{document}